\documentclass[10pt, conference, compsocconf]{IEEEtran}
\IEEEoverridecommandlockouts
\usepackage{etex}
\usepackage{cite}
\usepackage{amsmath,amssymb,amsfonts,amsthm}
\usepackage{graphicx}
\usepackage{textcomp}
\usepackage{xcolor}

\usepackage{booktabs}
\usepackage{threeparttable}

\usepackage{wasysym}

\usepackage[ruled,vlined,linesnumbered]{algorithm2e}

\usepackage[lambda,advantage,operators,sets,adversary,landau,probability,notions,logic,ff,mm,primitives,events,complexity,asymptotics,keys]{cryptocode}

\ifCLASSOPTIONcompsoc
  \usepackage[nocompress]{cite}
  \usepackage[caption=false,font=footnotesize,labelfont=sf,textfont=sf]{subfig}
\else
  \usepackage{cite}
  \usepackage[caption=false,font=footnotesize]{subfig}
\fi 

\def\BibTeX{{\rm B\kern-.05em{\sc i\kern-.025em b}\kern-.08em
    T\kern-.1667em\lower.7ex\hbox{E}\kern-.125emX}}
\begin{document}

\title{CryptoNN: Training Neural Networks over Encrypted Data
}

\author{
\IEEEauthorblockN{Runhua Xu, James B.D. Joshi, and Chao Li}
\IEEEauthorblockA{
\textit{School of Computing and Information, University of Pittsburgh}\\
runhua.xu@pitt.edu, jjoshi@pitt.edu, chl205@pitt.edu}
}

\maketitle

\begin{abstract}
Emerging neural networks based machine learning techniques such as deep learning and its variants have shown tremendous potential in many application domains.
However, they raise serious privacy concerns due to the risk of leakage of highly privacy-sensitive data when data collected from users is used to train neural network models to support predictive tasks.
To tackle such serious privacy concerns, several privacy-preserving approaches have been proposed in the literature that use either secure multi-party computation (SMC) or homomorphic encryption (HE) as the underlying mechanisms.
However, neither of these cryptographic approaches provides an efficient solution towards constructing a privacy-preserving machine learning model, as well as supporting both the training and inference phases.

To tackle the above issue, we propose a CryptoNN framework that supports training a neural network model over encrypted data by using the emerging functional encryption scheme instead of SMC or HE. We also construct a functional encryption scheme for basic arithmetic computation to support the requirement of the proposed CryptoNN framework.
We present performance evaluation and security analysis of the underlying crypto scheme and show through our experiments that CryptoNN achieves accuracy that is similar to those of the baseline neural network models on the MNIST dataset.
\end{abstract}


\section{Introduction}
Emerging deep learning (DL) architectures such as convolutional neural networks and recurrent neural networks show a huge promise for artificial intelligence (AI) based applications. 
Neural networks based machine learning techniques have succeeded in many fields including computer vision, speech/audio recognition, etc. \cite{lecun2015deep}.
Training of such a neural network model requires a huge amount of data to enable reliable predictive analytics.
Commercial AI service providers such as Google, Microsoft, and IBM have devoted a lot of efforts to build DL models for various intelligence applications, where the models are trained based on data collected from their customers.
Even though the AI-based applications bring convenience to daily life, it also raises serious privacy concerns due to the risk of leakage of highly privacy-sensitive data. Such privacy concerns are hindering the applicability of neural network models for real world applications \cite{bost2015machine, gilad2016cryptonets, mirhoseini2016cryptoml, rouhani2018deepsecure}.

To tackle the serious privacy concerns when applying neural network models in the cloud-based applications where users' privacy-sensitive data such as electronic health/medical records, location information, etc., are stored and processed, a well designed privacy-preserving neural networks framework is very essential so that the cloud and the data owners are not required to reveal their models and sensitive data, respectively. 
Existing privacy-preserving machine learning approaches are essentially of two types : (\romannumeral1) non-crypto based approaches such as SecureML \cite{mohassel2017secureml}, DeepSecure\cite{rouhani2018deepsecure}, and approaches proposed in \cite{mirhoseini2016cryptoml, shokri2015privacy, abadi2016deep}; and (\romannumeral2) crypto based approaches such as CryptoNets\cite{gilad2016cryptonets}, and approaches proposed in \cite{graepel2012ml, gonzalez2018supervised, wiesberg2018unsupervised, hesamifard2017cryptodl, chabanne2017privacy, park2018efficient, jiang2018secure}.
\tablename $\;$\ref{tab:comparsion_approach} illustrates the differences among the existing solutions.
Non-crypto approaches have several limitations such as significant communication overhead in the secure protocol based solutions where a high degree of coordination and information exchange are required \cite{mohassel2017secureml, rouhani2018deepsecure}, training overhead for the data owner \cite{mohassel2017secureml}, and utility loss \cite{abadi2016deep}.
Most of the existing crypto based approaches, such as  \cite{gilad2016cryptonets, graepel2012ml, gonzalez2018supervised, wiesberg2018unsupervised, hesamifard2017cryptodl, chabanne2017privacy, park2018efficient, jiang2018secure}, employ homomorphic encryption (HE) to protect the data, where the computation over encrypted data is allowed.
As the sensitive data encrypted by using HE can only be decrypted by the data owner, such proposed HE-based approaches, however, only support prediction over encrypted data using existing trained models, rather than training a model over encrypted data. This is because the computed results of HE is confidential to the server, and hence cannot be used for evaluation with the label during the back-propagation phase.
That is, the machine learning model should be trained on the plaintext data, and then the trained model can be applied over encrypted data to do the prediction.

\begin{table*}[!t]
    \centering
    \caption{Comparison of privacy-preserving approaches in machine learning models}
    \begin{threeparttable}
    \begin{tabular}{lcccll}
    \toprule
        Proposed Work    & Training   & Prediction  & Privacy \tnote{$\triangleright$}  & ML Model &  Approach    \\
    \midrule
        \cite{mirhoseini2016cryptoml}    & $\bullet$ & $\bullet$ & $\Circle$ & General  & Delegation\tnote{$\ddagger$} \\
        \cite{shokri2015privacy}    & $\bullet$ & $\circ$ & $\Circle$ & Deep Learning & Distributed\tnote{$\ast$}\\
        \cite{abadi2016deep} & $\bullet$ & $\circ$ & $\Circle$ & Deep Learning & Differential Privacy\tnote{$\diamond$}  \\
        SecureML\cite{mohassel2017secureml} & $\bullet$ &$\bullet$ & $\RIGHTcircle$ & General & Secure Protocol (SMC) \\
        DeepSecure\cite{rouhani2018deepsecure}  & $\bullet$ &$\bullet$ & $\RIGHTcircle$ & Deep Learning & Secure Protocol (Garbled Circuits) \\
        CryptoNets\cite{gilad2016cryptonets}, \cite{graepel2012ml}, \cite{gonzalez2018supervised}, \cite{wiesberg2018unsupervised},\cite{ hesamifard2017cryptodl}, \cite{chabanne2017privacy}, \cite{park2018efficient}, \cite{jiang2018secure} & $\circ$ & $\bullet$ & $\CIRCLE$ & Covers All & Homomorphic Encryption (HE)   \\
        \cite{bost2015machine} & $\bullet$ &$\bullet$ & $\CIRCLE$ & Limited ML\tnote{$\dagger$} & HE + Secure Protocol \\
        CryptoNN (our work) & $\bullet$ &$\bullet$ & $\CIRCLE$ & Neural Networks & Functional Encryption \\
    \bottomrule
    \end{tabular}
    \begin{tablenotes}
        \footnotesize
        \item[$\triangleright$] This column indicates the privacy strength/guarantee such as mild approach $\Circle$ (e.g. differential privacy) and strong guarantee $\CIRCLE$ (e.g. crypto system). 
        \item[$\dagger$] It only supports Hyperplane Decision, Naïve Bayes, and Decision Trees models.
        \item[$\ddagger$] The data owner trains the model by itself and outsources partial computation in a privacy-preserving setting.
        \item[$\ast$] The model is trained in a distributed manner where each data owner trains a partial model on their private data. 
        \item[$\diamond$] It applies differential privacy method on the training data.
    \end{tablenotes}
    \end{threeparttable}
    \label{tab:comparsion_approach}
\end{table*}

Consider a computer aided diagnostic application scenario: several distributed federal clinics want to train an AI model to help diagnose the images generated from magnetic resonance imaging (MRI), X-ray, etc.
The limited IT infrastructure or AI experts in such clinics means that they cannot train the model by themselves, and hence have to rely on a third party service, such as a cloud based machine learning service.
However, the regulations for protecting patients' healthcare records require a secure approach for training machine learning models on the privacy-sensitive data collected from different federal clinics.
To the best of our knowledge, only the work proposed in \cite{bost2015machine} supports both training and predictive analysis over encrypted data by integrating several crypto schemes ( i.e., Quadratic Residuosity cryptosystem, Paillier cryptosystem, and homomorphic encryption) with secure protocols designed for them.
However, it only supports limited types of machine learning model as presented in \tablename$\;$\ref{tab:comparsion_approach}.

To tackle the challenge of training a model over encrypted data in a simpler manner, in this paper, we propose a novel framework called CryptoNN, where a neural network model is trained over encrypted data in a very simple way without the overhead of interactive communication protocols (i.e., without complex secure protocols), while also supporting predictive analytics in a privacy-preserving way.
The underlying crypto scheme in our proposed CryptoNN is the emerging \emph{functional encryption} \cite{boneh2011functional}.
Unlike traditional encryption schemes such as homomorphic encryption, where decryption reveals all or nothing, in a \textit{functional encryption} scheme, the decryption keys can also reveal some partial information about the plaintext\cite{boneh2012functional}. 
To be concrete, for a function $f(\cdot)$, an authority holding a master secret key can generate a key $\sk_{f}$ that enables the computation of the function $f_{\sk_{f}}(\cdot)$ on the encrypted data. 
For instance, using function related key $\sk_{f}$ the decryptor can compute $f_{\sk_{f}}(x)$ from the encrypted data, $\text{enc(}x\text{)}$, without revealing the plaintext $x$.
CryptoNN can compute several permitted functions over the privacy-sensitive data that is encrypted by a distributed set of data owners, and only acquires the computed results instead of the original plaintext data.
This makes it possible to train a neural network model over encrypted data without the overhead of the communication protocols.
The proposed CryptoNN framework relies on a secure matrix computation scheme which is based on existing functional encryption for the inner-product scheme proposed by Abdalla et al. in \cite{abdalla2015simple} and our newly constructed functional encryption for the basic arithmetic operations. 

We summarize our key contributions as follows:
\begin{itemize}
    \item We propose a functional encryption scheme for basic arithmetic operations to support basic computation over encrypted data.
    \item We also propose a secure matrix computation scheme, where the elements are protected by those functional encryption schemes.
    \item We propose a general CryptoNN framework built upon the secure computation components to support the training of neural network models over encrypted data. Essentially, CryptoNN inserts a \textit{secure feed-forward} step and a \textit{secure back-propagation/evaluation} step into the training phase of the normal neural network.
    Also, we present a crypto-based \textit{convolutional neural network} model as an instantiation of the CryptoNN scheme.
    \item Finally, we analyze the security of the proposed work, and present experiments to evaluate the performance of the underlying secure matrix computation and the CryptoNN schemes. We show that the accuracy of CryptoNN is similar to that of the original NN model.
\end{itemize}
To the best of our knowledge, this is the first approach for training a model over encrypted data using the functional encryption scheme.

\noindent\textbf{Organization}. 
The paper is organized as follows. 
In Section \ref{sec:background}, we introduce the background and preliminaries.
We propose our CryptoNN framework and its related secure components in Section \ref{sec:cryptonn}.
The security analysis and experimental evaluation are presented in Section \ref{sec:evaluation}.
We discuss the related work in Section \ref{sec:related_work} and conclude the paper in Section \ref{sec:conclusion}.

\section{Background and Preliminaries}
\label{sec:background}

\subsection{Application Scenario}

Most of the existing AI-based applications such as Google's Cloud AI and Microsoft's Azure Machine Learning are built using the cloud-client or server-client architecture, where the client holds the data and the cloud/server is responsible for training a machine learning model for a specific task such as classification, regression, etc., and/or providing a predictive analytics service for the client. 

In this paper, we consider the same architecture, but with stronger privacy guarantees as all the data owned by a client is in an encrypted form.
To be specific, in the training phase, what the client needs to do is encrypting its sensitive data as required and then sending it to the server. 
The server computes several permitted functions over encrypted data, acquires the computation result, and builds the final neural network model. 
Similarly, the data to be used for prediction is also encrypted in the prediction phase, and then the server outputs the prediction using the trained model. 

Note that we put more efforts in the training phase in this paper, which is currently the biggest challenge in crypto-based machine learning models \cite{bost2015machine}.
In the prediction phase, it is still possible to apply (\romannumeral1) existing homomorphic encryption (HE) based approaches (e.g., CryptoNets\cite{gilad2016cryptonets}, and approaches in \cite{graepel2012ml, gonzalez2018supervised, wiesberg2018unsupervised, hesamifard2017cryptodl, chabanne2017privacy, park2018efficient, jiang2018secure}), or (\romannumeral2) our proposed functional encryption (FE) based approach. 
Neither our FE based approach nor the existing HE based approaches allow the server to learn the prediction data.
The only difference is that an HE based approach does not allow the server to learn the prediction result, while the FE based approach allows that.
Such a setting provides flexible choices for the client with varying levels of privacy concerns.

\subsection{Functional Encryption}
\label{sec:background:feip}
Recently proposed approaches such as those in \cite{goldwasser2014multi, boneh2015semantically, waters2015punctured, garg2016candidate, carmer20175gen, lewi20165gen}, have focused their attention on the theoretical feasibility or existence of functional encryption, and thus constructing schemes for general functionalities at the expense of efficiency.
In particular, those schemes rely on strong primitives such as indistinguishable obfuscation or multilinear maps that are prohibitively inefficient \cite{kim2018function}.
As a result, these schemes are far from being practically deployable. 

In our paper, we adopt the functional encryption approach proposed in \cite{abdalla2015simple} as the basis for the matrix computation over encrypted data.
It is a simple functional encryption scheme only supporting \textit{inner-product} computation, whose security is based on DDH assumption.
Even though the functionality is limited to \textit{inner-product} computation, it is the state-of-the-art practical construction so far.
To support the element-wise computation (i.e., product, division, addition, and subtraction), we also propose our construction based on the same DDH security assumption in the next section.

\begin{table}[t]
    \centering
	\begin{threeparttable}
	\caption{General notations and symbols in this paper}
	\vspace{-2mm}
	\label{table:notation}
	\begin{tabular}{ll}
		\toprule
		Symbols 	& Description       \\
		\midrule
		\textit{mpk}    & The public key generated by the authority.\\
		\textit{msk}    & The master secret key generated by the authority.\\
		$f(\cdot)$      & The general function over arbitrary input.\\
		$\textit{sk}_{f}$& The secret key related to function $f(\cdot)$.\\
		$[\eta]$        & A set $\{1, 2, ..., \eta\}$. \\
		$\textbf{x}$    & The bold lowercase letters denoting vectors. \\
		$\textbf{X}$    & The bold uppercase letters denoting matrices. \\
		$[[d]]$         & The cihpertext of $d$. \\ 
		$\partial$      & The partial derivative symbol. \\ 
		\bottomrule
	\end{tabular}
	\end{threeparttable}
	\vspace{-5mm}
\end{table}

Here, we present the specific construction of functional encryption for an inner product proposed in \cite{abdalla2015simple}.
Suppose the inner-product functionality is defined as follows: 
$$f(\textbf{x},\textbf{y}) = \langle\textbf{x},\textbf{y}\rangle = \sum^{\eta}_{i=1}(x_{i}y_{i}),$$
where $\eta$ is the length of the vectors $\textbf{x}$ and $\textbf{y}$.
Several general symbols and notations are presented in \tablename$\;$\ref{table:notation}.
The functional encryption scheme for the inner-product functionality $f(\textbf{x},\textbf{y})$ is defined as \texttt{FEIP} = (\texttt{Setup, KeyDerive, Encrypt, Decrypt}), as follows:
\begin{itemize}
    \item $\texttt{Setup(}1^{\lambda}, 1^{\eta}\texttt{)}$: The algorithm first generates two samples as
    $(\GG, p, g)$ $\sample$ $\texttt{GroupGen}(1^{\lambda})$, and $\textbf{s}$ $=$ $(s_1$ $,...,$ $s_{\eta})$ $\sample$ $\ZZ^{\eta}_{p}$ on the inputs of security parameters $\lambda$ and $\eta$,
    and then sets $\textit{mpk} = (g, h_{i} = g^{s_{i}})_{i\in [\eta]}$ and $\textit{msk}$ $=$ $\textbf{s}$.
    It returns the pair (\textit{mpk, msk}).
    \item $\texttt{KeyDerive(}\textit{msk}, \textbf{y}\texttt{)}$: The algorithm outputs the function secret key $\textit{sk}_{f}=\langle\textbf{y},\textbf{s}\rangle$ on the inputs of master secret key \textit{msk} and vector \textbf{y}.
    \item $\texttt{Encrypt(}\textit{mpk},\textbf{x}\texttt{)}$: The algorithm first chooses a random $r\sample \ZZ_{p}$ and computes $\textit{ct}_0 = g^{r}$. For each $i \in [\eta]$, it computes $\textit{ct}_i = h_i^r\cdot g^{x_{i}}$.
    Then the algorithm outputs the ciphertext $\textit{ct} = (\textit{ct}_0, (\textit{ct}_i)_{i \in [\eta]})$.
    \item $\texttt{Decrypt(}\textit{mpk}, \textit{ct}, \textit{sk}_{f}, \textbf{y}\texttt{)}$: The algorithm takes the ciphertext \textit{ct}, the public key \textit{mpk} and functional key $\textit{sk}_{f}$ for the vector \textbf{y}, and returns the discrete logarithm in basis $g$, i.e., $g^{\langle\textbf{x},\textbf{y}\rangle} = \prod_{i\in[\eta]}\textit{ct}_{i}^{y_{i}}/ \textit{ct}_{0}^{\textit{sk}_{f}}$.
\end{itemize}
Note that even though it is impossible to recover the exponent of generator $g^r$ in the group $\GG$ according to the DDH assumption, it is still possible to recover the inner-product $\langle\textbf{x},\textbf{y}\rangle$ in the discrete logarithm with the same basis, $g$, as $\langle\textbf{x},\textbf{y}\rangle \ll r$.
There exists efficient approaches to computing the discrete logarithm such as \textit{baby-step giant-step} algorithm \cite{terr2000modification}.
We refer the readers to \cite{abdalla2015simple} for the correctness proof of the decryption.

\subsection{Neural Networks}
A neural network usually refers to learning a model that is a hierarchical and non-linear in structure consisting of several layers, where each layer includes several neural units (i.e., artificial neurons). In a neural network, each layer receives the data generated by its previous layer and outputs the processed data for the next layer.
In particular, the raw data is encoded properly and fed into the first layer of a neural network, also known as the input layer. 
Then, these features from the raw data are gradually mapped to higher-level abstractions via the iterative update (a.k.a, feed-forward and back-propagation) in the intermediate (hidden) layers of the neural network until the convergence condition is achieved (e.g., the specified number of iteration).
These mapping abstractions known as learned neural network model then can be used to predict the label in the last layer (i.e., the output layer).

With the emergence of the deep neural network (a.k.a deep learning) techniques, several architectures have been proposed that have been successfully used in different application domains.
For instance, the convolutional neural network (CNN, or ConvNet) \cite{krizhevsky2012imagenet} is a kind of deep learning model that is typically applied towards learning the features in the computer vision domain.
A typical CNN architecture includes the following layer: the input layer, the convolutional layer, the pooling layer (also known as the down-sampling layer), activation layer (e.g., ReLU layer, sigmoid layer, etc.), fully connected layer, and the output layer.

In this paper, we present a framework that is able to train a neural network model over encrypted data.
Here, we present several common processing operations that are computed in each layer or neural unit.
\begin{itemize}
    \item \textit{Dot-product function}: It is used in the general hidden layer, where the dot-product operation needs to be executed between the trained parameters of a specific hidden layer and the outputs of the previous layer. 
    \item \textit{Weighted-sum function}: It is used in the convolutional layer, where the filter or kernel (i.e., the layer's parameter) is convolved across the height and width of the input volume by computing each weighted-sum of the input and the filter matrix. 
    \item \textit{Pooling function}: It is used in the pooling layer where the maximal/average value of some of the components of the feeding layer is computed.
    \item \textit{Activation function}: It is used in the activation layer, where the computation is based on a specific activation function such as rectified linear (ReLu) function, sigmoid function, tanh function, etc. 
    \item \textit{Cost function}: It is used in the output layer, where the output of the neural network model is measured by a specified cost function with ground truth label to evaluate the performance of the model. 
\end{itemize}

\section{CryptoNN Framework}
\label{sec:cryptonn}

\subsection{Overview}

\begin{figure*}[!t]
  \centering\includegraphics[scale=0.5]{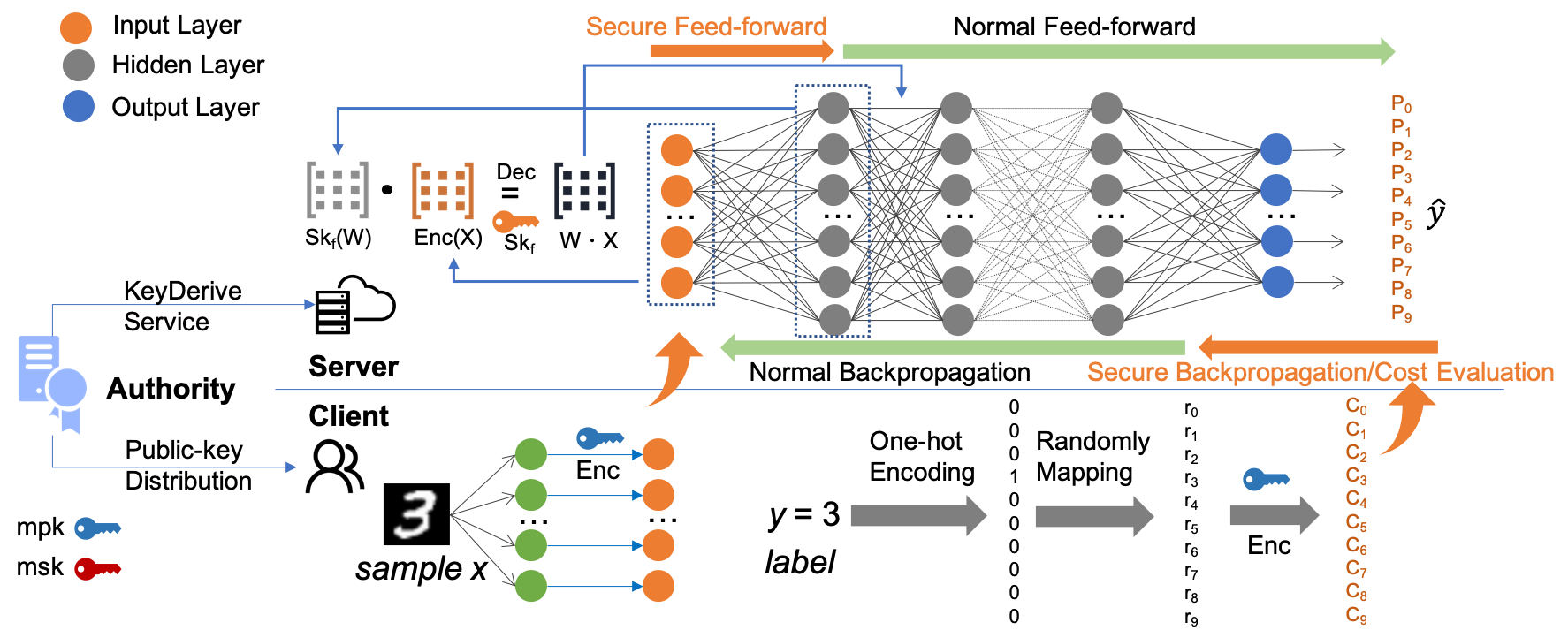}
  \vspace{-3mm}
  \caption{An overview of the CryptoNN framework. Note that for the demonstration purpose, we just use one image sample to illustrate the secure process of the CryptoNN framework, as well as the one-hot encoding method for the label. Besides, as our CryptoNN framework only focuses on the data processing which takes place between the input layer and first hidden layer in the secure feed-forward step, and between the output layer and last hidden layer in the secure back-propagation/evaluation step, we omit the details of the neural network in the middle of hidden layers.}
  \vspace{-5mm}
  \label{fig:cryptonn_framework}
\end{figure*}

The proposed CryptoNN framework includes three entities: \textit{authority}, \textit{server}, and \textit{client} as depicted in the \figurename \ref{fig:cryptonn_framework}.

The \textit{authority} is responsible for setting up the crypto parameters such as public key \textit{mpk} and master secret key \textit{msk}. 
Then it holds the \textit{msk} and distributes the public keys to the servers and clients.
Besides, it also generates the function derived key for the \textit{server} to decrypt the function result.

The \textit{client} first needs to pre-process the training data as required, and then encrypt all the pre-processed training data using the public key, \textit{mpk}.
The pre-processing operation can be quite simple. For instance, in the computer vision related neural network model, the pretreatment involves mapping each image into a vector. 
Suppose that the data is a color image with size $28\times28$. 
The client needs to map all $28\times28\times3$ pixel values into one vector and then encrypt the vector.
Similar pretreatment and encryption are applied to the label.
Once the encrypted training data is ready, the client sends it to the server for training a neural network model.

The \textit{server} collects all the encrypted data from a distributed set of clients.
For each training iteration, the \textit{server} first needs to acquire the function derived key, $\textit{sk}_{f}$, from the \textit{authority} for a specific function (e.g., dot-product, element-wise subtract, etc.) representing the underlying computation between the input layer and the first hidden layer; then it needs to decrypt the function result and continue the process in the rest of the hidden layers.
For instance, in the the feed-forward phase, a typical process for the first hidden layer is $a=g(\textbf{W}\textbf{X} + \textbf{b}),$ where $a$ is the output of the layer, $g$ is the activation function, and $\textbf{b}$ is the bias. $\textbf{X}$ represents the features of samples from the input layer, and $\textbf{W}$ is the parameters of each neural node in the first hidden layer.

In our CryptoNN framework, the input $\textit{enc}(\textbf{X})$ is protected by functional encryption scheme; thus, it is not feasible to compute $\textbf{W}\textbf{X}$ directly.
Thus, the process needs to be transferred to 
$$
a = g(\textit{sk}_{f}(\textbf{W})\cdot\text{enc}(\textbf{X}) + \textbf{b}) = g(f(\textbf{W}\textbf{X}) + \textbf{b}).
$$
Even though the \textit{server} does not learn the value of \textbf{X}, it can compute the result (i.e., $f(\textbf{W}\textbf{X})$) using the function derived key, $sk_{f}$.
As a result, the \textit{server} can continue the feed-forward process in the neural networks until it reaches the output layer, where similar secure computation is applied to the encrypted label.
Note that the label \textbf{Y} also should be pre-processed before encryption to prevent a \textit{direct inference attack} discussed later in Section \ref{sec:cryptonn:febo}.
For instance, to prevent inference, the label should be mapped to a random number first.
More details will be discussed in Section \ref{sec:eval:security}.

\noindent\textbf{\textit{Distributed data source}}. 
Our CryptoNN framework does not limit each neural network model to correspond to only one client.
Actually, the model can be trained over multiple, distributed data sources, where the only requirement is that the training data should be encrypted using the same public key.

\noindent\textbf{\textit{Comparison with existing solutions}}.
As our CryptoNN framework is a crypto approach to deal with privacy issues in a machine learning model, we compare the existing homomorphic encryption (HE) based solutions here rather than secure multi-party computation based approaches, which will be discussed in the related work section later.

Intuitively, both the HE-based solutions and our CryptoNN support the computation over encrypted data, and hence, they do not reveal the data to the server. 
The main difference is that the result of the computation using HE is still in encrypted form, while the result of the computation using CryptoNN is plaintext.
Such a difference allows CryptoNN to do secure computation in the training phases, while other HE-based models only support prediction over encrypted data.
Besides, the intermediate data is not ciphertext in the hidden layers, which intuitively indicates that CryptoNN will be more efficient than the HE-based solutions.

In the rest of this section, we first present the proposed underlying crypto scheme, namely, \textit{ functional encryption}; then we present the construction of the secure matrix computation based on this functional encryption crypto-system.
Finally, we present our CryptNN framework that is built using the proposed secure components, where a general neural network model can be transferred to CryptoNN to enable training over encrypted data.

\subsection{Functional Encryption for Basic Operations}
\label{sec:cryptonn:febo}
Even though the functional encryption for inner-product can be employed to compute the dot-product over encrypted matrix, there is still lack of functional encryption schemes to support element-wise computation. 
Here, we propose our construction of the functional encryption scheme supporting basic operations, i.e., addition, subtraction, multiplication, and division.
Suppose the basic functionality is defined as follows:
$$ f_{\varDelta\in[+,-,*,/]}(x,y) = x \varDelta y ,$$
where $\varDelta$ denotes the specific computation between $x$ and $y$.
The functional encryption scheme for the basic functionality $f_{\varDelta}(x,y)$ is defined as \texttt{FEBO} = \texttt{(Setup, KeyDerive, Encrypt, Decrypt)}.
Each algorithm is defined as follows:
\begin{itemize}
    \item \texttt{Setup(}$1^{\lambda}$\texttt{)}: It takes the security parameter $\lambda$ as the input, and outputs the key pair (\textit{mpk, msk}).
    It first generates the group samples $(\GG, p, g)$ $\sample$ $\texttt{GroupGen}(1^{\lambda})$ and randomly chooses the secret $s$ $\sample$ $\ZZ_{p}$.
    Then, it sets \textit{msk} $= s$ and \textit{mpk} $= (h, g)$, where $h = g^{s}$.
    \item \texttt{KeyDerive(}\textit{mpk}, \textit{msk}, \textit{cmt}, $\varDelta$, $y$\texttt{)}: The algorithm takes the master secert key, \textit{msk}, the commitment, \textit{cmt}, the input of function $y$ as the inputs and outputs the secret key, $\textit{sk}_{f_{\varDelta}}$.
    The algorithm has different key generation approaches according to different arithmetic computations. The function derived key  $\textit{sk}_{f_{\varDelta}}$ is generated as follows:
    $$ 
    \textit{sk}_{f_{\varDelta}} = \left\{ 
    \begin{array}{ll} 
    \textit{cmt}^{s}\cdot g^{-y}& \text{if } \varDelta = +,\\ 
    \textit{cmt}^{s}\cdot g^{y} & \text{if } \varDelta = -,\\
    (\textit{cmt}^{s})^{y} & \text{if } \varDelta = *,\\ 
    (\textit{cmt}^{s})^{y^{-1}} & \text{if } \varDelta = /.
    \end{array} \right.
    $$
    \item \texttt{Encrypt(}\textit{mpk}, $x$\texttt{)}: The algorithm takes the pubic key, \textit{mpk}, and the input of function $f_{\varDelta}$ as the inputs, and outputs a commitment, \textit{cmt}, and the ciphertext of $x$.
    To achieve that, it first randomly chooses a nonce, $r$ $\sample$ $\ZZ_{p}$, and generates the commitment \textit{cmt} $= g^{r}$.
    The ciphertext is computed as $\textit{ct}=h^{r}\cdot g^{x}$.
    \item \texttt{Decrypt(}\textit{mpk}, $\textit{sk}_{f_{\varDelta}}$, \textit{ct}, $\varDelta$, $y$ \texttt{)}: Based on the inputs - the public key \textit{mpk}, the function derived key $\textit{sk}_{f_{\varDelta}}$ for function $f_{\varDelta}$ with another input $y$, and the ciphertext \textit{ct} - the algorithm recovers the discrete logarithm in basis $g$ as follows:
    $$ 
    \textit{g}^{f_{\varDelta}(x,y)} = \left\{ 
    \begin{array}{ll} 
    ct/sk & \text{if } \varDelta = + \text{ or } -,\\
    (ct)^{y}/sk & \text{if } \varDelta = *,\\
    (ct)^{y^{-1}}/sk & \text{if } \varDelta = /.
    \end{array} \right.
    $$
\end{itemize}
As we explained in Section \ref{sec:background:feip}, it is practical to recover the exponent, i.e., the result of $f_{\varDelta}(x,y)$.

\noindent\textit{Correctness}. For all (\textit{mpk}, \textit{msk}) $\leftarrow$ \texttt{Setup(}$1^{\lambda}$\texttt{)}, all $x,y \in \ZZ_{p}, \varDelta \in [+,-,*,/]$, for $\textit{sk}_{f_{\varDelta}}$ $\leftarrow$ \texttt{KeyDerive(}\textit{msk}, \textit{cmt}, $\varDelta$, $y$\texttt{)}, and \textit{cmt},\textit{ct} $\leftarrow$ \texttt{Encrypt(}\textit{mpk}, $x$\texttt{)}, we have \texttt{Decrypt(}\textit{mpk}, $\textit{sk}_{f_{\varDelta}}$, \textit{ct}, $\varDelta$, $y$\texttt{)} computed as follows:
$$
\left\{ 
    \begin{array}{ll} 
    ct/sk = (h^rg^x)/((g^{r})^sg^{(\mp y)}) = g^{x\pm y} & \text{if } \varDelta = \pm,\\
    (ct)^{y}/sk = (h^rg^x)^y/(g^r)^{sy} = g^{xy}& \text{if } \varDelta = *,\\
    (ct)^{y^{-1}}/sk = (h^rg^x)^(y^{-1})/(g^r)^{sy^{-1}} = g^{x/y} & \text{if } \varDelta = /.
    \end{array} \right.
$$

\noindent\textit{Remark}. 
The construction of \texttt{FEBO} is derived from ElGamal encryption \cite{elgamal1985public}.
The security proof of \texttt{FEBO} scheme is presented in Section \ref{sec:eval:security}.
The \texttt{FEBO} scheme can resist an attacker (i.e., unauthorized user without function key) trying to break the encrypted data, but cannot prevent the \textit{direct inference attack}, where the attacker (i.e., authorized decryptor with function key) can infer the encrypted data $\text{Enc}(x)$ using the final function result $f(x\Delta y)$ and its own $y$ in the four basic arithmetic computations.
The issue will be addressed at the construction of framework level components.

\subsection{Secure Matrix Computation}
\label{sec:cryptonn:smc}
Matrix computation is the basic operation in the training and prediction phases of neural networks.
To support matrix computation over encrypted data, we construct an encrypted matrix computation method, called \textit{secure matrix computation} in the rest of the paper, by employing the functional encryption for inner-product proposed in \cite{abdalla2015simple} and our proposed functional encryption for the basic operations presented in Section \ref{sec:cryptonn:febo}.

Suppose that we need to compute a function $f \in \mathcal{F}$, where $\mathcal{F}$ is the permitted set of functions over encrypted matrix \textbf{X} and \textbf{Y} that come from the \textit{client} and the \textit{server}, respectively.   
The permitted function set includes dot-product and element-wise arithmetic computation.
The specific scheme is presented in Algorithm \ref{alg:smc}.

\begin{algorithm}[t]
    \footnotesize
    \caption{secure matrix computation scheme}
    \label{alg:smc}
    \KwIn{$\textbf{X}$, $\textbf{Y}$, a permitted function $f \in \mathcal{F}$ over encrypted matrix.}
    \KwOut{Computation result $\textbf{Z}$ and pre-process result [[\textbf{x}]], [[\textbf{X}]], $\textbf{sk}_{f}$. }
    \SetKwProg{Fn}{function}{}{}
    The authority initializes and distributes public key $\textit{mpk}_\text{FEIP}, \textit{mpk}_\text{FEBO}$ \\ 
    \Fn{secure-computation([[\textbf{x}]], [[\textbf{X}]], f, $\mathcal{F}$, $\textbf{sk}_{f,\textbf{y}/\textbf{Y}}$, $\textbf{Y}$)}{
        initialize an empty matrix $\textbf{Z}$ \\
        \If{$ f \in \mathcal{F}$ \textbf{and} $f$ is dot-product}{
            \For{$i\leftarrow 0$ \KwTo $\text{size of}$ $\textbf{sk}_{f,\textbf{y}}$}{
                $\textbf{y}$ := $i$-th row of $\textbf{Y}$ \\
                \For{$j\leftarrow 0$ \KwTo $\text{size of}$ [[\textbf{x}]]}{
                    \textbf{Z}[i][j] :=  \texttt{FEIP.Decrypt(}$\textit{mpk}_\text{FEIP}$,\textbf{x}[j],$\textbf{sk}_{f,\textbf{y}}$[i],\textbf{y}\textit{)}
                }
            }
        }
        \Else{
            \For{$i\leftarrow 0$ \KwTo row of [[\textbf{X}]]}{
                \For{$j\leftarrow 0$ \KwTo column of [[\textbf{X}]]}{
                    \textbf{Z}[i][j] :=  \texttt{FEBO.Decrypt(}$\textit{mpk}_\text{FEBO}$,$\textbf{sk}_{f,\textbf{Y}}$[i][j], \lbrack[\textbf{X}]][i][j],$\mathcal{F}$,\textbf{Y}[i][j]\textit{)}
                }
            }
        }
        \Return $\textbf{Z}$
    }
    \Fn{pre-process-encryption($\textbf{X}$)}{
        initialize an empty list [[\textbf{x}]] and an empty matrix [[\textbf{X}]]\\
        \For{$i\leftarrow 0$ \KwTo column size of \textbf{X}}{
            $\textbf{x}$ := $i$-th column of $\textbf{X}$ \\
            $\lbrack$[\textbf{x}]$\rbrack$[i] := \texttt{FEIP.Encrypt(}$\textit{mpk}_\text{FEIP}$,$\textbf{x}$\textit{)} \\
            \For{$j \leftarrow 0$ \KwTo row size of \textbf{X}}{
                [[\textbf{X}]][j][i] := \texttt{FEBO.Encrypt(}$\textit{mpk}_\text{FEBO}$,\textbf{X}[j][i]\textit{)} \\
            }
        }
        \Return [[\textbf{x}]], [[\textbf{X}]]
    }
    \Fn{pre-process-key-derivative($\textbf{Y}, f, \mathcal{F}$)}{
        initialize an empty list $\textbf{sk}_{f,\textbf{y}}$ and matrix $\textbf{sk}_{f,\textbf{Y}}$\\
        \For{$i\leftarrow 0$ \KwTo row size of \textbf{Y}}{
            \If{$ f \in \mathcal{F}$ \textbf{and} $f$ is dot-product}{
                $\textbf{y}$ := $i$-th row of $\textbf{Y}$ \\
                $\textbf{sk}_{f,\textbf{y}}$[i] :=  $\textit{sk}_{f}$ from authority by $\textbf{y}$ \\
            }
            \Else{
                \For{$j \leftarrow 0$ \KwTo column size of \textbf{Y}}{
                    $\textbf{sk}_{f,\textbf{Y}}$[i][j] :=  $\textit{sk}_{f}$ from authority by \textbf{Y}[i][j] \\
                }
            }
        }
        \Return $\textbf{sk}_{f,\textbf{y}}$ or $\textbf{sk}_{f,\textbf{Y}}$
    }
\end{algorithm}

The scheme includes three parts: \textit{pre-process-encryption} function, \textit{pre-process-key-derivative} function, and \textit{secure-computation} function.
As a client, it just needs to encrypt the matrix using the \textit{pre-process-encrypt} function and send out the ciphertext (lines 14-21).
For the server, it first needs to decide the specific function $f \in \mathcal{F}$, and then request/prepare the function derived key $\textbf{sk}_{f}$ from the \textit{authority} (lines 25-30).
When the encrypted data from the clients and the function derived key are ready, the \textit{server} starts the processing over encrypted data using \textit{secure-computation} function as described in lines 2-13.

Note that even though the secure dot-product computation can also be achieved using secure element-wise multiplication method in our proposed scheme, we still separate it as an independent function here due to efficiency considerations.
The related analysis is presented in Section \ref{sec:evaluation}.
For simplicity, we just use the 2-dimensional array (i.e., matrix) to demonstrate the secure computations in the neural network model. 
Such a design is similar to the setting in the \textit{NumPy}, which is the most widely used library for the scientific computing in the machine learning area.
Note that the underlying decryption involves time-consuming operations in the secure computation scheme; however, our scheme also supports parallel computation.
For instance, it just needs to make decryption code segment (line 8 and line 12) parallelized.
Our experiments show that the parallelization will increase the efficiency significantly (see Section \ref{sec:evaluation} for more details).

\subsection{CryptoNN Framework}

\begin{algorithm}[t]
    \footnotesize
    \caption{The CryptoNN framework}
    \label{alg:cryptonn}
    \KwIn{The encrypted training data [[\textbf{x}]] and label [[\textbf{Y}]], the layers information $\mathcal{L}$, the hyperparameters $\mathcal{H}$, the permitted function set $\mathcal{F}$ over encrypted data.}
    \KwOut{the trained parameters $\mathcal{P}$.}
    \SetKwProg{Fn}{function}{}{}
    \Fn{CryptoNN( [[\textbf{x}]], [[\textbf{Y}]], $\mathcal{L}$, $\mathcal{H}$, $\mathcal{F}$)}{
        $\mathcal{P} \leftarrow$ initialize parameters according to $\mathcal{L}$ \\
        \ForEach{iterator in total iterations}{
            $\textbf{sk}_{\mathcal{F}, \mathcal{P}^{[1]}} \leftarrow$ \textit{pre-process-key-derive}($\mathcal{P}^{[1]}$, $\mathcal{F}$) \\
            $\mathcal{R}_{X} \leftarrow$ \textit{secure-computation}([[\textbf{x}]], $\mathcal{F}$, $\textbf{sk}_{\mathcal{F},\mathcal{P}^{[1]}}$, $\mathcal{P}^{[1]}$) \\
            $\mathcal{A} \leftarrow $ normal feed-forward process using $\mathcal{R}_{X}$, $\mathcal{P}$\\
            $\textbf{sk}_{\mathcal{F}, \mathcal{A}^{[l]}} \leftarrow$ \textit{pre-process-key-derive}($\mathcal{A}^{[l]}$, $\mathcal{F}$) \\
            $\mathcal{R}_{\mathcal{A}^{[l]}} \leftarrow$ \textit{secure-computation}([[\textbf{Y}]], $\mathcal{F}$, $\textbf{sk}_{\mathcal{F},\mathcal{A}^{[l]}}$, $\mathcal{A}^{[l]}$) \\
            cost $\mathcal{C} \leftarrow$ cost evaluation using $\mathcal{R}_{\mathcal{A}^{[l]}}, \mathcal{A}$ on the cost function \\
            grads $\mathcal{G} \leftarrow$ normal back-propagation using $\mathcal{R}_{\mathcal{A}^{[l]}}, \mathcal{A}, \mathcal{H}$ \\
            $\mathcal{P} \leftarrow$ update parameters using existing $\mathcal{P}$, and $\mathcal{G}, \mathcal{H}$ \\
        }
        \Return $\mathcal{P}$
    }
\end{algorithm}

The proposed CryptoNN framework is built on the secure matrix computation scheme and its underlying functional encryption schemes.
As shown in Algorithm \ref{alg:cryptonn}, we present the general CryptoNN.
As the training data is encrypted, the CryptoNN framework inserts two rounds of secure computations presented in Section \ref{sec:cryptonn:smc} to deal with the computation between the parameters and the encrypted data (i.e., lines 4-5 and lines 7-8) for each iteration.
Such two rounds of secure computation only take place at the beginning of feed-forward processing and back-propagation processing, and hence, they are called \textit{\textbf{secure feed-forward}} and \textit{\textbf{secure back-propagation and evaluation}}, respectively, in the rest of the paper (see \figurename \ref{fig:cryptonn_framework} as the example).

\noindent\textbf{Scalability}. CryptoNN is adaptive to any existing neural network model if the \textit{secure feed-forward} and \textit{secure back-propagation / evaluation} steps in such models can be narrowed down to process the basic matrix computation supported by the permitted function set $\mathcal{F}$.
As we cannot enumerate all the existing neural network models to explain the general applicability of our proposed CryptoNN, here, we first use a simple neural network model for a binary classification task for illustration.
Then, we present a concrete case: applying the proposed CryptoNN framework to the typical Convolutional Neural Network (CNN) in the next section.

Suppose the \textit{secure feed-forward} step of the model computes 
$$
\textbf{A}=\theta(\textbf{Z})= \theta(\textbf{W}\textbf{X} + \textbf{b}),
$$
where $\theta(\cdot)$ is a sigmoid function $\theta(z) = \frac{1}{1+e^{-z}}$ and \textbf{A} is the output of the layer.
The process in the \textit{secure back-propagation/evaluation} step of the model includes the sigmoid function for the output layer, and the cost function is 
$$
E = \frac{1}{2}\sum_{i}(\hat{y}^{(i)} - y^{(i)})^{2}
,$$
where $\hat{y}^{(i)}$ is the prediction and $y^{(i)}$ is the label.
The \textit{secure feed-forward} process can be split into two steps: (\romannumeral1) computing $\textbf{W}\textbf{X}$ first, which is a dot-product supported by the secure matrix computation scheme; and then (\romannumeral2) computing the remaining parts of the formula.
For \textit{secure evaluation}, it can compute $\hat{\textbf{Y}} - \textbf{Y}$ first, which is element-wise subtraction also supported by CryptoNN.
To update the parameter for \textit{secure back-propagation} we need to compute $\frac{\partial E}{\partial \textbf{W}}$ according to gradient descent method. 
Based on the partial derivative rule, we know that 
$$\frac{\partial E}{\partial \textbf{W}} = \frac{\partial \textbf{E}}{\partial \textbf{A}} \cdot \frac{\partial \textbf{A}}{\partial \textbf{Z}} \cdot \frac{\partial \textbf{Z}}{\partial \textbf{W}},$$
such that, 
$$\frac{\partial \textbf{Z}}{\partial \textbf{W}} = \textbf{A}^{[l-1]}, \frac{\partial \textbf{A}}{\partial \textbf{Z}} = \textbf{A}^{[l]}(1-\textbf{A}^{[l]}), \frac{\partial E}{\partial \textbf{A}} = \textbf{A}^{[l]} - \textbf{Y}, $$
where $\textbf{A}^{[l-1]}$ is the output of the last hidden layer, and $\hat{\textbf{Y}} = \textbf{A}^{[l]}$ is the output (prediction) of the output layer.
Obliviously, the computation related to label \textbf{Y} is the basic element-wise subtraction that is also supported by CryptoNN.

\noindent\textbf{Prediction}. We do not present the details of the prediction (or inference) phase of the CryptoNN framework here, as the prediction can be viewed as the sub-process of the training phases. 
To be specific, the prediction phase only covers the \textit{secure feed-forward} and normal feed-forward, and gets the output of the neural network model.
As the trained model of CryptoNN is also plaintext, it can also be integrated with existing homomorphic encryption-based solutions at the prediction phase.
Hence, it provides a flexible choice of privacy setting, namely, keeping the predicted label confidential or not.
If the user prefers the confidential predicted label, the HE-based prediction is employed; otherwise, the FE-based prediction can be applied.

\subsection{A Concrete Case: CryptoCNN}
\label{sec:cryptonn:cnn}
Here, we use another concrete and relatively more complicated neural network model, namely, LeNet-5, a classic convolutional neural network \cite{lecun1998gradient} for multiple classification, to illustrate the scalability of CryptoNN.

Except for the input and output layer, the LeNet-5 model includes five hidden layers: convolutional layer (C1), the average pooling layer (S2), the convolutional layer (C3), the average pooling layer (S4) and fully connected layer (C5).
As our CryptoNN only focuses on the \textit{secure feed-forward} and \textit{secure back-propagation / evaluation} steps in the model, we need to focus on the C1 and the output layers.
To be concrete, we need to address the \textit{padding} and \textit{convolution} operations in the \textit{secure feed-forward} step, and \textit{softmax output} function, \textit{softmax cross-entropy loss} function in the \textit{secure back-propagation/evaluation} step.

\begin{figure}[!t]
  \centering\includegraphics[scale=0.45]{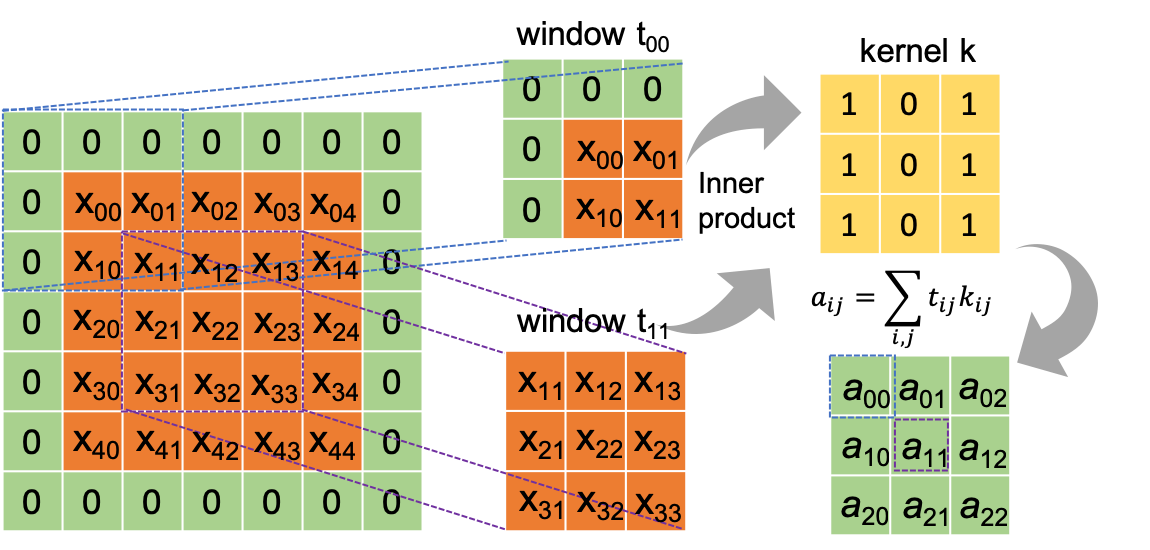}
  \vspace{-6mm}
  \caption{An illustration of padding and convolution operations in the CNN model. Suppose that the original image size is $5\times5\times1$, and the padding is 1. The filter size is $3\times3\times1$. The stride is 2, hence, the output size is $3\times3\times1$.}
  \vspace{-3mm}
  \label{fig:cryptonn_conv}
\end{figure}

\subsubsection{\textit{secure feed-forward} step}
It takes place in the first convolutional layer.
As depicted in \figurename \ref{fig:cryptonn_conv}, we illustrate classic \textit{padding} and \textit{convolution} operations.
Typically, the original image is surrounded by \textit{zero-padding} data if the padding is required in the CNN model.
Next, a sliding window with the same size of the filter (a.k.a. kernel) is set to capture parts of the image, and then next step is to do the convolution with the filter to fill the corresponding element in the output volume.
In particular, the computation is the sum of element-wise product of each corresponding element in the two matrix, namely, $a_{ij} = \sum_{i,j}(t_{ij}k_{ij})$, where $t_{ij}$ and $k_{ij}$ are the elements from the window and filter volume, respectively.
The sliding window will keep moving according to the stride length.

In the CryptoNN framework, the original image (i.e., training data) is pre-processed under the protection of functional encryption.
Thus, the padded image will be a ``mixed'' matrix, as depicted in \figurename \ref{fig:cryptonn_conv}, where the yellow element is the encrypted data, while the green element is the plaintext \textit{zero-padding} data.
When doing the convolution, there are two types of sliding windows: fully encrypted window and partially encrypted window, e.g., $t_{11}$ and $t_{00}$, respectively, as shown in \figurename \ref{fig:cryptonn_conv}.
To tackle such an issue, we present the \textit{secure convolution computation} scheme, as shown in Algorithm \ref{alg:scc}.
In essence, it is similar to the secure matrix computation scheme, as the convolution operation of two matrices can be converted to the computation of dot-product of two vectors.
As the architecture is fixed in the adopted CNN model, the \textit{client} needs to learn the padding strategy and the filter size from the \textit{server}, and then pre-process the training data according to the filter size (lines 9-16).
The \textit{server} first prepares the key $\textbf{sk}_{f, \textbf{K}}$ based on the filter \textbf{K} (lines 17-20), and then executes the secure convolution operation for the corresponding filter/kernel (lines 2-8).
Note that we only present the \textit{one-filter} case in the algorithm, it is obviously applicable to \textit{multi-filter} case.

\begin{algorithm}[t]
    \footnotesize
    \caption{secure convolution scheme}
    \label{alg:scc}
    \KwIn{training data \textbf{X}, filter $\textbf{K}$, a permitted function $f \in\mathcal{F}$ over encrypted matrix, public parameter $\textit{mpk}_{\texttt{FEIP}}$}
    \KwOut{the result $\textbf{Z}$. }
    \SetKwProg{Fn}{function}{}{}
    Server prepare the $\textbf{sk}_{f, \textbf{K}}$ for element-wise product for filter \textbf{K}\\ 
    \Fn{secure-convolution(\textbf{T}, $\textbf{sk}_{f, \textbf{K}}$, $\textbf{K}$)}{
        initialize an empty matrix $\textbf{Z}$ \\
        \ForEach{element [[\textbf{t}]] in \textbf{T}}{
            \textbf{k} $\leftarrow$ transfer \textbf{K} to vector \\
            t := \texttt{FEIP.Decrypt(}$\textit{mpk}_\text{FEIP}$, $\textbf{sk}_{f,\textbf{k}}$, [[\textbf{t}]], f, $\mathcal{F}$, \textbf{k}\textit{)} \\
            fill t into \textbf{Z} at right position\\
        }
        \Return $\textbf{Z}$
    }
    \Fn{pre-process-encryption(\textbf{X})}{
        initialize an empty window list \textbf{T} \\
        $\textbf{X}^{'} \leftarrow $ \textbf{X} with padding \\
        \Repeat{slide window finished}{
            \textbf{t} $\leftarrow $ abstract window matrix from $\textbf{X}^{'}$ and transfer to vector \\
            append \texttt{FEIP.Encrypt(}$\textit{mpk}_\text{FEIP}$, $\textbf{t}$\textit{)} into \textbf{T}\\
        }
        \Return \textbf{T}
    }
    \Fn{pre-process-key-derivative(\textbf{K}, $f, \mathcal{F}$)}{
        \textbf{k} $\leftarrow$ transfer \textbf{K} to vector \\
        $\textbf{sk}_{f,\textbf{K}}$ := request key from authority by \textbf{k} \\
        \Return $\textbf{sk}_{f,\textbf{K}}$
    }
\end{algorithm}

\subsubsection{\textit{secure back-propagation/evaluation} step}
It takes place at the output layer.
We assume that the output is the \textit{softmax} function, namely, 
$p_{i} = \frac{e^{a_{i}}}{\sum_{k=1}^{N}{e^{a_{k}}}}$, where $p_i$ denotes the probability of data $x$ belonging to the class $i$, and $N$ denotes the total number of categories.
Here, $a_i$ is the element of the output vector $\textbf{a}$ of the fully connected layer for data $x$.
Also, suppose that the label $y$ is encoded as a vector \textbf{y} using the one-hot method.
Thus, the \textit{cross-entropy loss} is 
$$ L = - \sum_{i=1}^{N}{y_i\text{log}p_{i}},$$ where $y_i$ is the component of label \textbf{y} and $p_i$ is the component of output vector \textbf{p} of the softmax function.

In the CryptoNN framework, \textbf{y} is encrypted using functional encryption. However, obliviously, the loss function $L = - \langle\textbf{y},\textbf{p}^{'}\rangle$ is a kind of inner-product computation that is supported by  CryptoNN, where $\textbf{p}^{'}$ is the logarithmic vector of \textbf{p}.

To do \textit{secure back-propagation} for the parameter update, it needs to compute $$\frac{\partial L}{\partial a_j} = \frac{\partial L}{\partial p_i}\frac{\partial p_i}{\partial a_j}.$$
Briefly, we can calculate the partial derivative of $\frac{\partial p_i}{\partial a_i}$ first and the result is 
$$
\frac{\partial p_i}{\partial a_j} = \left\{ 
\begin{array}{rl} 
p_i(1-p_j)  &\text{if } i = j,\\
- p_jp_i    &\text{if } i \ne j. 
\end{array} \right.
$$
Then, $\frac{\partial L}{\partial a_j} = p_i - y_i$, and the vectorized expression for all labels is 
$\frac{\partial L}{\partial \textbf{A}} = \textbf{P} - \textbf{Y}.$
The computation related to label \textbf{Y} is the basic element-wise subtraction that is also supported by CryptoNN.

\section{Implementation and Evaluation}
\label{sec:evaluation}

\subsection{Security Analysis}
\label{sec:eval:security}

\noindent\textit{The Discrete Diffie-Hellman assumption}.
Suppose that the group triplet ($\GG, p, g$) is generated by a PPT algorithm with the security parameter $1^{\lambda}$, where $\GG$ is the group of a $\lambda$-bit prime order $p$, and $g$ is the generator.
The Discrete Diffie-Hellman assumption states that the tuples ($g, g^a, g^b, g^{ab}$) are computationally indistinguishable from ($g, g^a, g^b, g^c$), where $\{a, b, c\}\in\ZZ_{p}$ are chosen independently and uniformly at random.

Theorem \ref{theo:security_theorem} establishes the security of FEBO scheme. 
\newtheorem{theorem}{\textbf{Theorem}}
\begin{theorem}
    \label{theo:security_theorem}
    \textit{If the Discrete Diffie-Hellman problem is hard (i.e., DDH assumption holds), the constructed FEBO scheme has selective security against chosen-plaintext attacks (IND-CPA).}
\end{theorem}

\begin{proof}
    For the sake of contradiction, assume that there exists an adversary $\adv$ that can break the IND-CPA security of FEBO scheme with non-negligible advantage. 
    
    Given the adversary $\adv$ that runs in time $t$ and has advantage  $\epsilon$, we can construct an adversary $\bdv$ for the DDH problem that runs in time $t+O(1)$ and also has the advantage $\epsilon$.
    Then $\bdv (g, h_1=g^a, h_2=g^r, h_3=g^c)$ is constructed as follows: (i) set the \textit{mpk} as $(g, h_1)$, $b \sample \{0,1\}$, and ciphertext as $(h_2, h_3\cdot g^{x\Delta y})$; (ii) run $\adv$(\textit{mpk},$c$) to get the output $b^{'}$; (iii) if $b=b^{'}$, $\bdv$ guesses it as the valid DDH tuple; otherwise, $\bdv$ guesses it is the random DDH tuple.
    If $c=ra$, then the ciphertext is a legitimate encryption of $g^{x\Delta y}$.
    If $c$ is uniformly distributed, independent of $a$ and $r$, then the ciphertext is just the original output. 
    This makes $\bdv$ a perfect simulator of a random oracle in this case.
    Thus, $\bdv$ also has the advantage $\epsilon$ to break the DDH assumption, which violates our security assumption.
    Thus, the adversary $\adv$ does not have the non-negligible advantage to break the FEBO scheme.
\end{proof}

As the FEIP scheme has been employed in the work proposed in \cite{abdalla2015simple}, where the security proof has been presented, we do not provide related security analysis here.

\noindent\textit{Security analysis of CryptoNN}.
The underlying crypto scheme of CryptoNN is the functional encryption scheme, for which the security has been explained above. 
Here, we present the security analysis of CryptoNN at a high level.
The \textit{authority} is the third party trusted authority and is supposed to be not in collusion with anyone.
From the perspective of  the \textit{server}, it can only acquire the mapping relationship between the training data and the label, where the original training/label data collected from multiple data sources is protected by the functional encryption.
Even for the simple label that has high similarity, for instance, one label maps to a set of training data, the encrypted result is uniformly distributed in the ciphertext space at random for each same label.
Thus, such mapping relationship does not reveal any private information.

The fact that the output of the \textit{secure feed-forward} and \textit{secure back-propagation} steps is the plaintext also indicates a trade-off problem.
On the one hand, it will increase the efficiency of training a neural network model.
On the other hand, the output of the first hidden layer may include the intermediate data that can be used to infer partial information.
A pre-processing approach such as a randomly mapping as mentioned in Section \ref{sec:cryptonn} can avoid the direct inference for the label, while the inner-product result cannot be used to directly infer the factor.
In addition, we also assume that the server is not an active attacker that will collect ``representative plaintext dataset'' for the encrypted training data.
Thus, the attack approaches as proposed in \cite{ligier2017information, carpov2018illuminating} is not applicable in our designed framework. 
Besides, the content related to complex inference attack such as measuring neural network memorization and extracting secrets is beyond the scope of this paper.
We refer the reader to \cite{carlini2018secret} for more details and corresponding countermeasures.

\begin{figure*}[t] 
    \centering 
    \subfloat[pre-processing for encryption]{
        \includegraphics[scale=0.28]{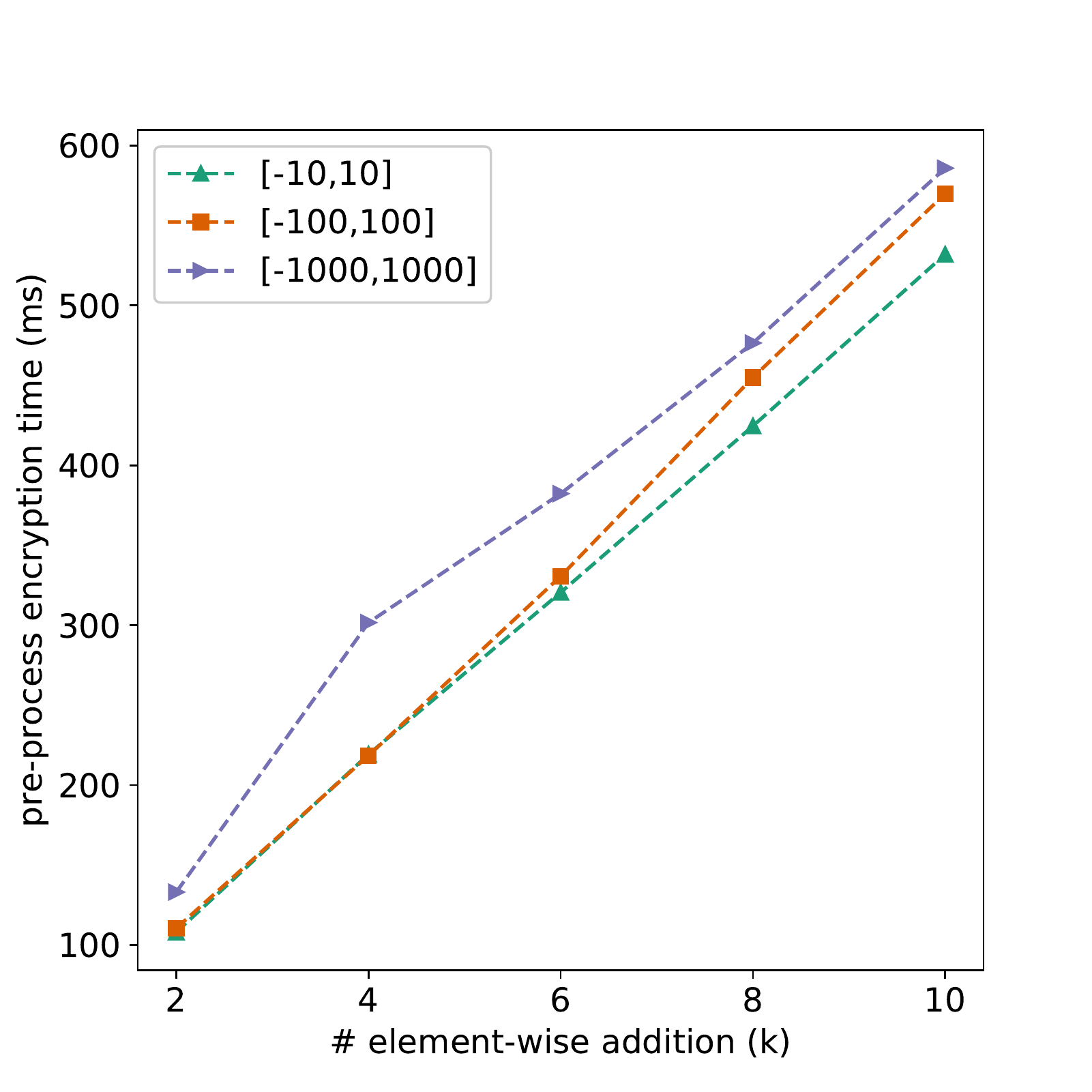}
        \label{fig:smc_add:enc}
    }
    \hspace{-5mm}
    \subfloat[pre-processing for function key]{
        \includegraphics[scale=0.28]{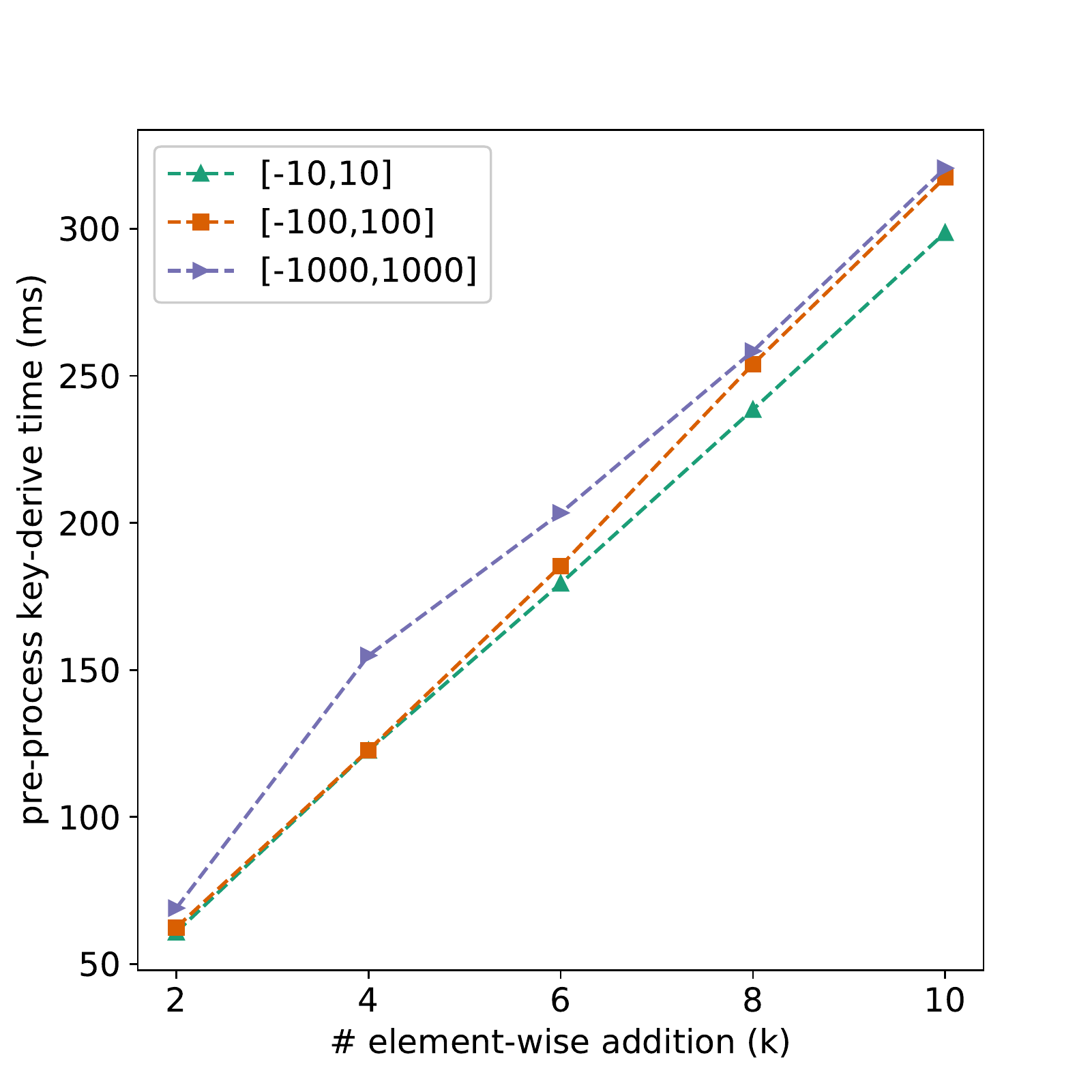}
        \label{fig:smc_add:key}
    }
    \hspace{-5mm}
    \subfloat[secure addition computation]{
        \includegraphics[scale=0.28]{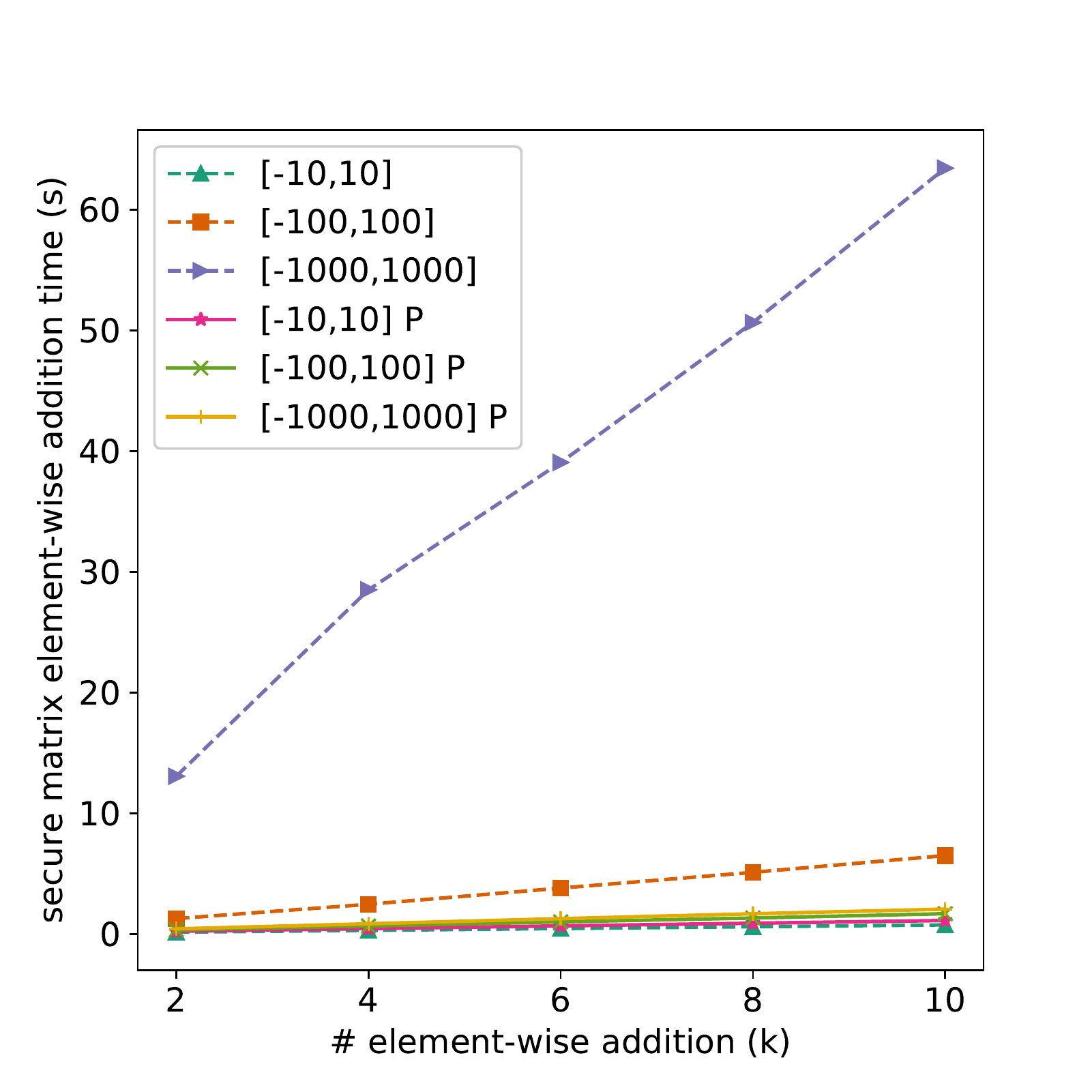}
        \label{fig:smc_add:dec}
    }
    \hspace{-5mm}
    \subfloat[secure addition (parallelized)]{
        \includegraphics[scale=0.28]{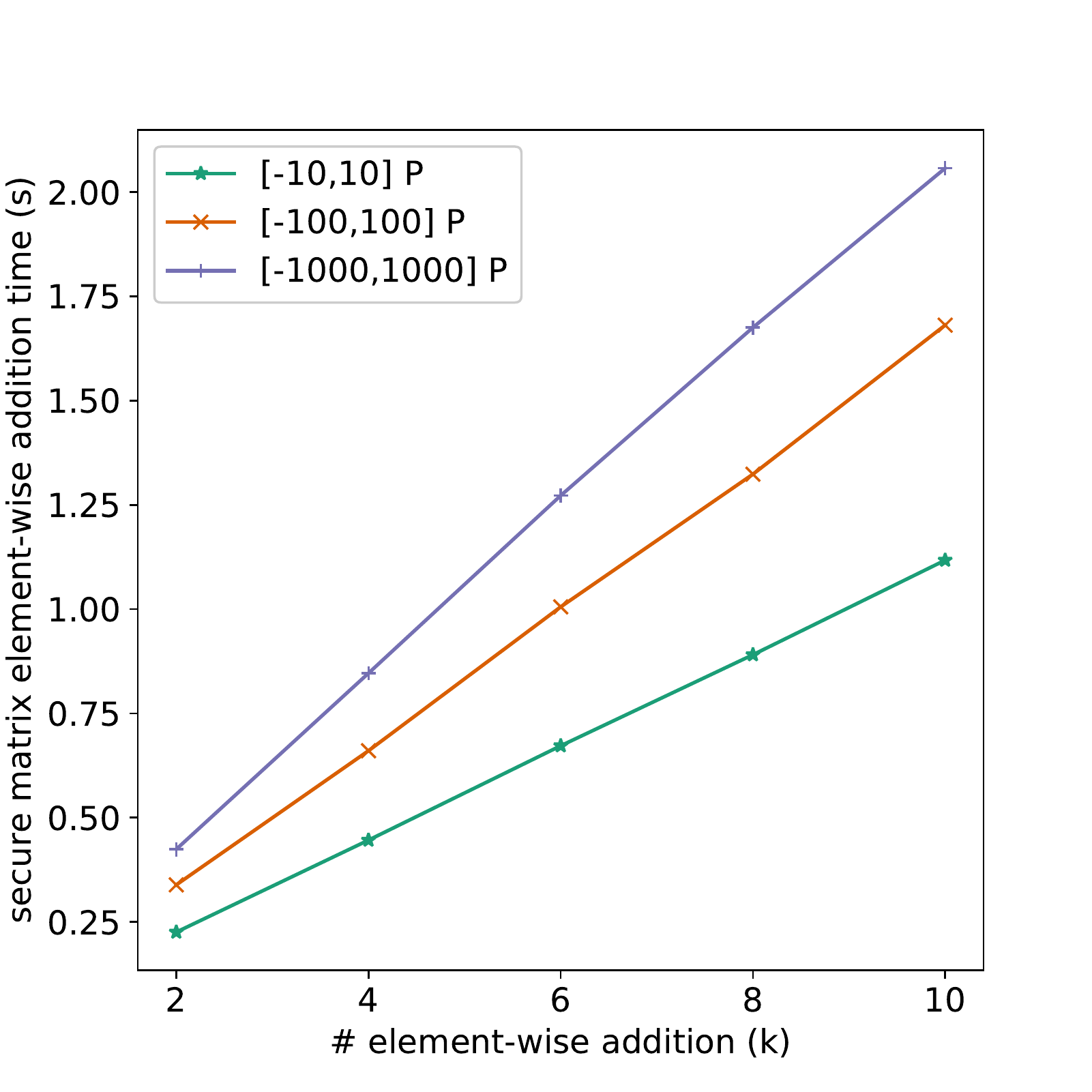}
        \label{fig:smc_add:dec_parallel}
    }
    \vspace{-1mm}
    \caption{The time cost of element-wise addition in secure matrix computation scheme.} 
    \vspace{-5mm}
    \label{fig:smc_add} 
\end{figure*}

\begin{figure*}[t] 
    \centering 
    \subfloat[pre-processing for encryption]{
        \includegraphics[scale=0.28]{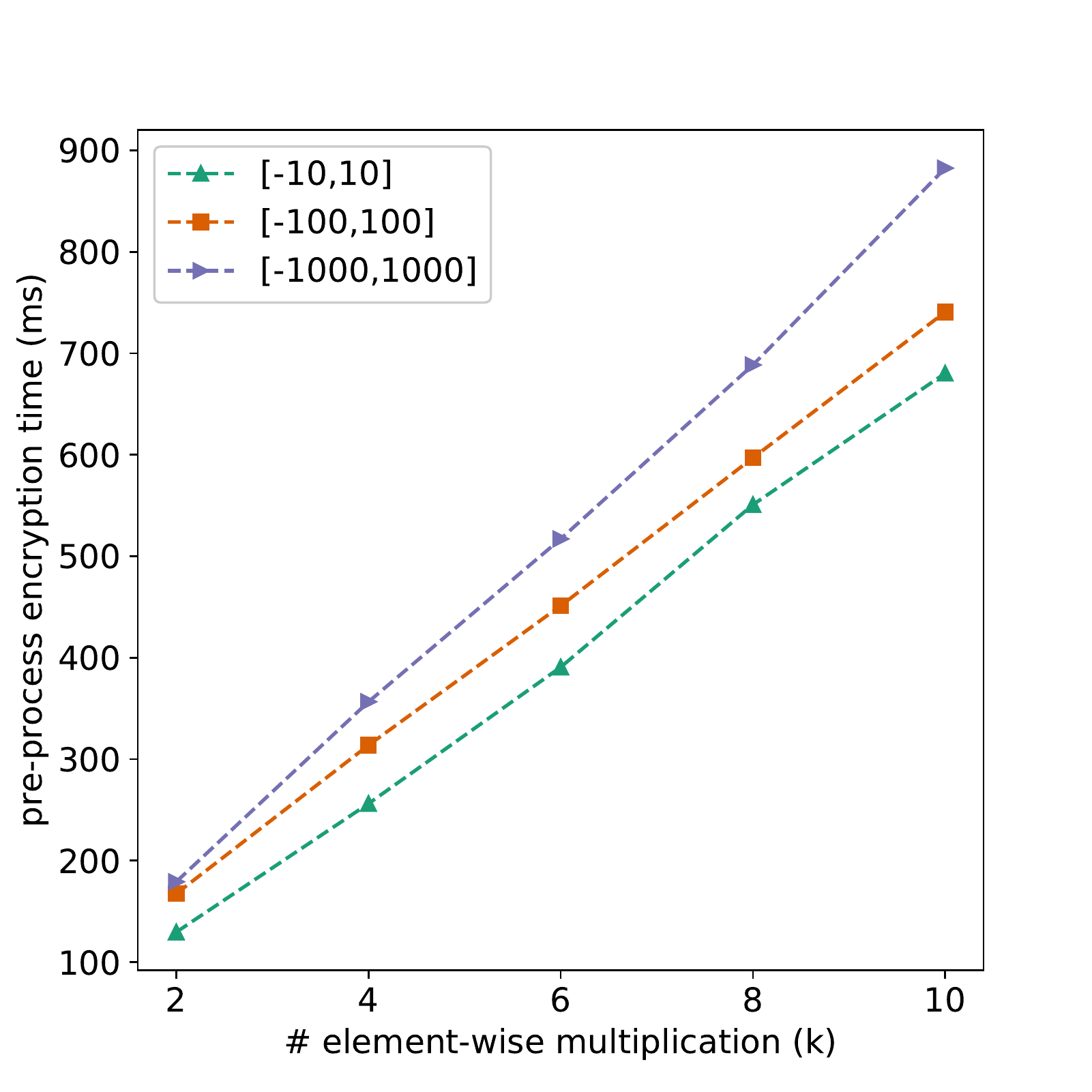}
        \label{fig:smc_mul:enc}
    }
    \hspace{-5mm}
    \subfloat[pre-processing for function key]{
        \includegraphics[scale=0.28]{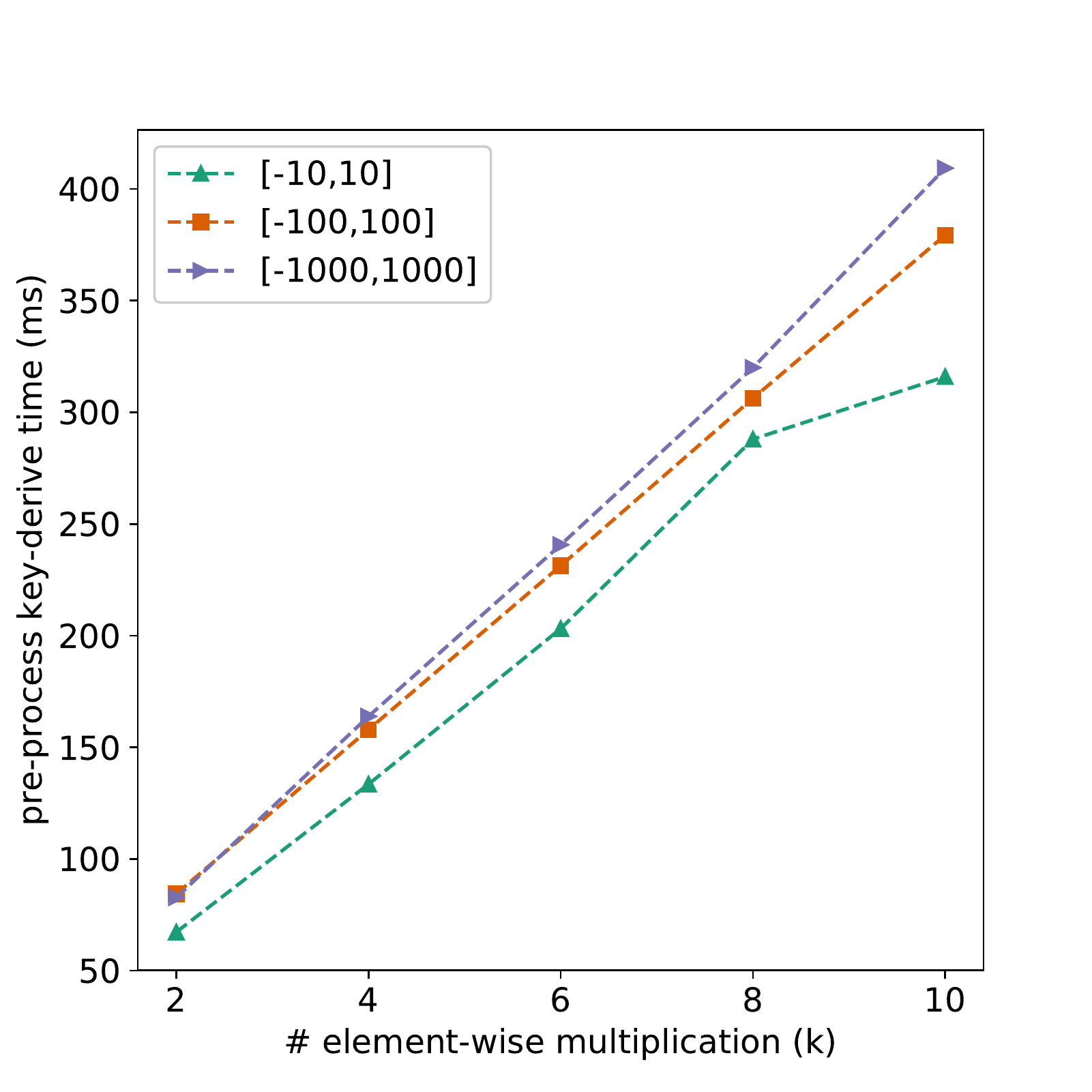}
        \label{fig:smc_mul:key}
    }
    \hspace{-5mm}
    \subfloat[secure multiplication computation]{
        \includegraphics[scale=0.28]{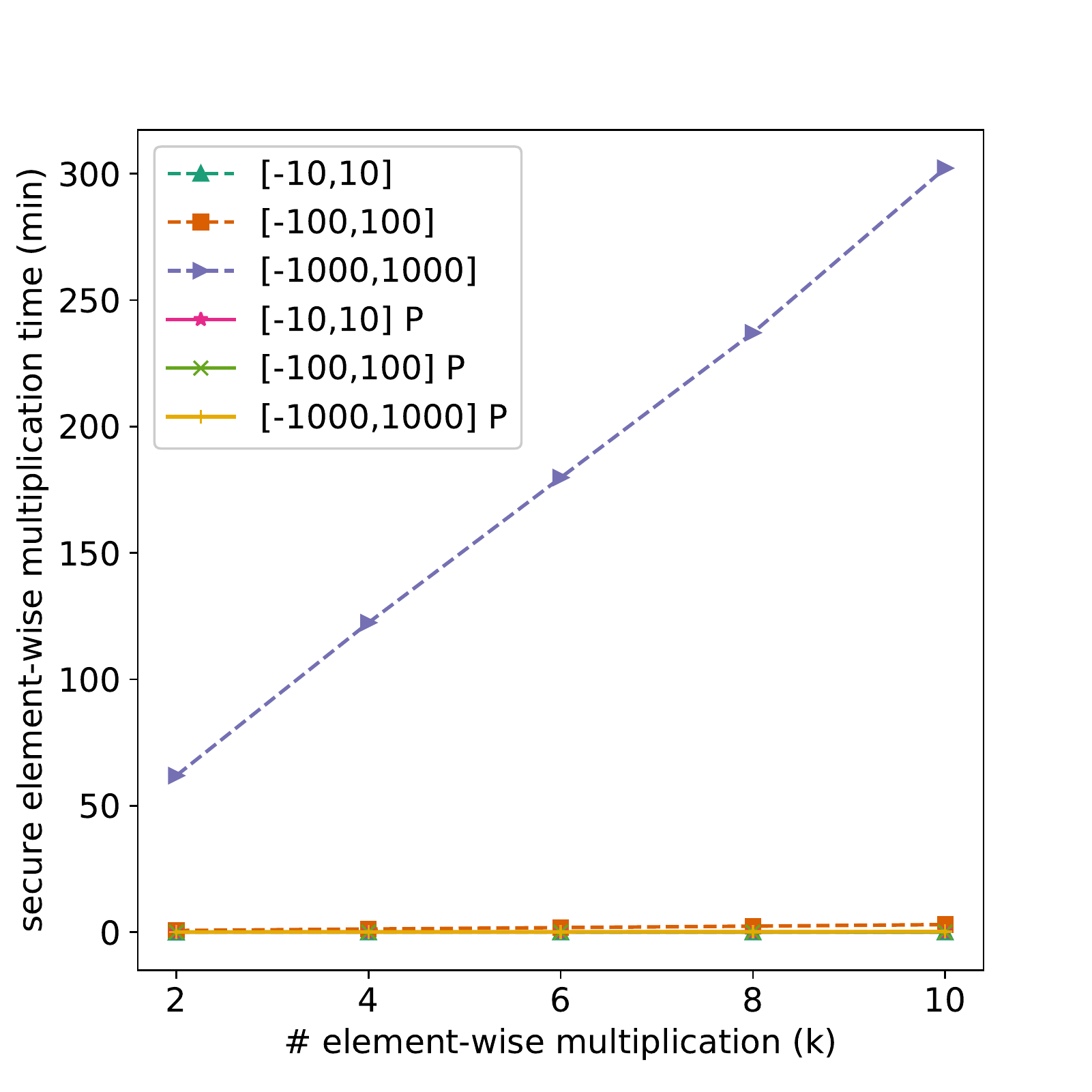}
        \label{fig:smc_mul:dec}
    }
    \hspace{-5mm}
    \subfloat[secure multiplication (parallelized)]{
        \includegraphics[scale=0.28]{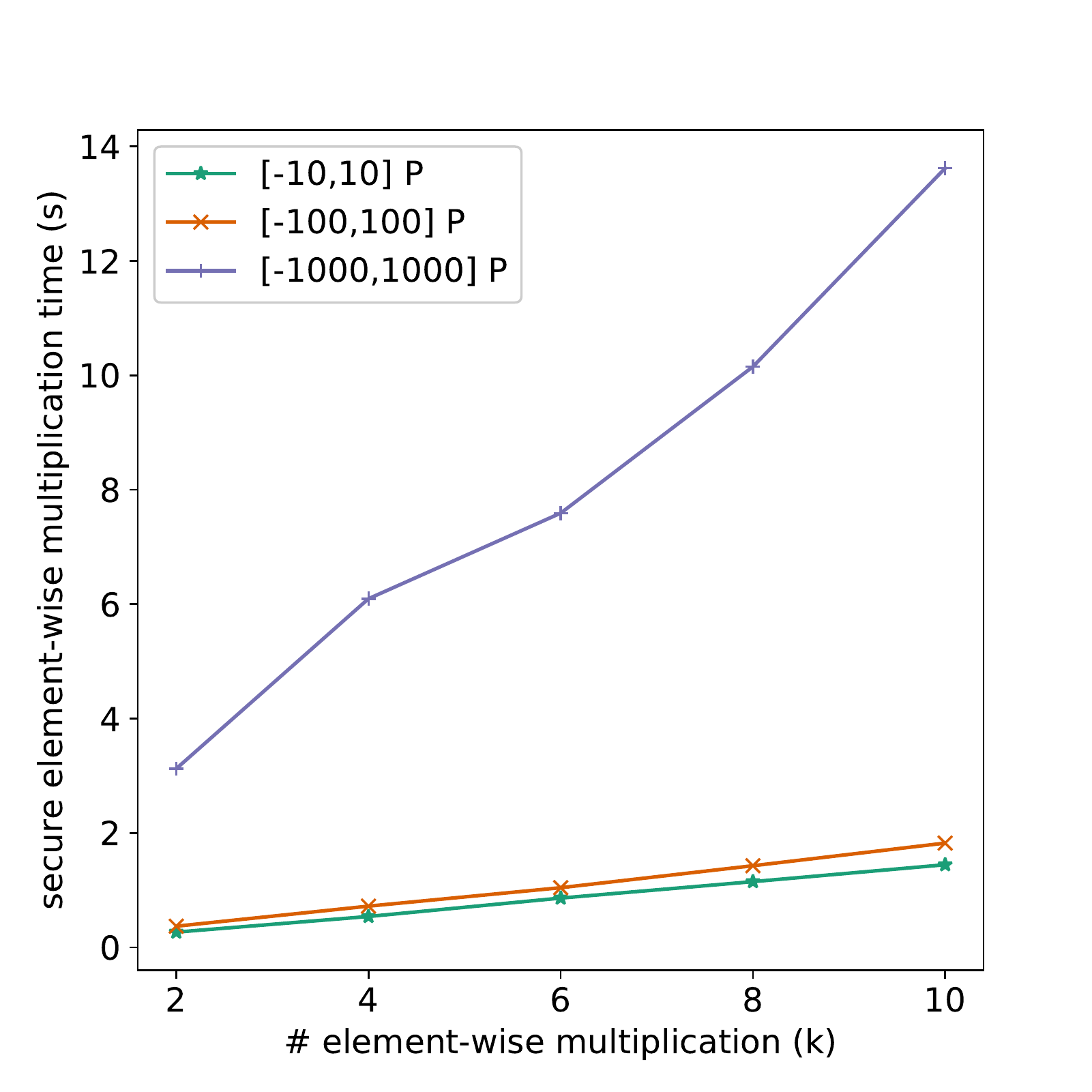}
        \label{fig:smc_mul:dec_parallel}
    }
    \vspace{-1mm}
    \caption{The time cost of element-wise multiplication in secure matrix computation scheme.}
    \vspace{-5mm}
    \label{fig:smc_mul} 
\end{figure*}
 
\begin{figure*}[t] 
    \centering 
    \subfloat[pre-processing for encryption]{
        \includegraphics[scale=0.28]{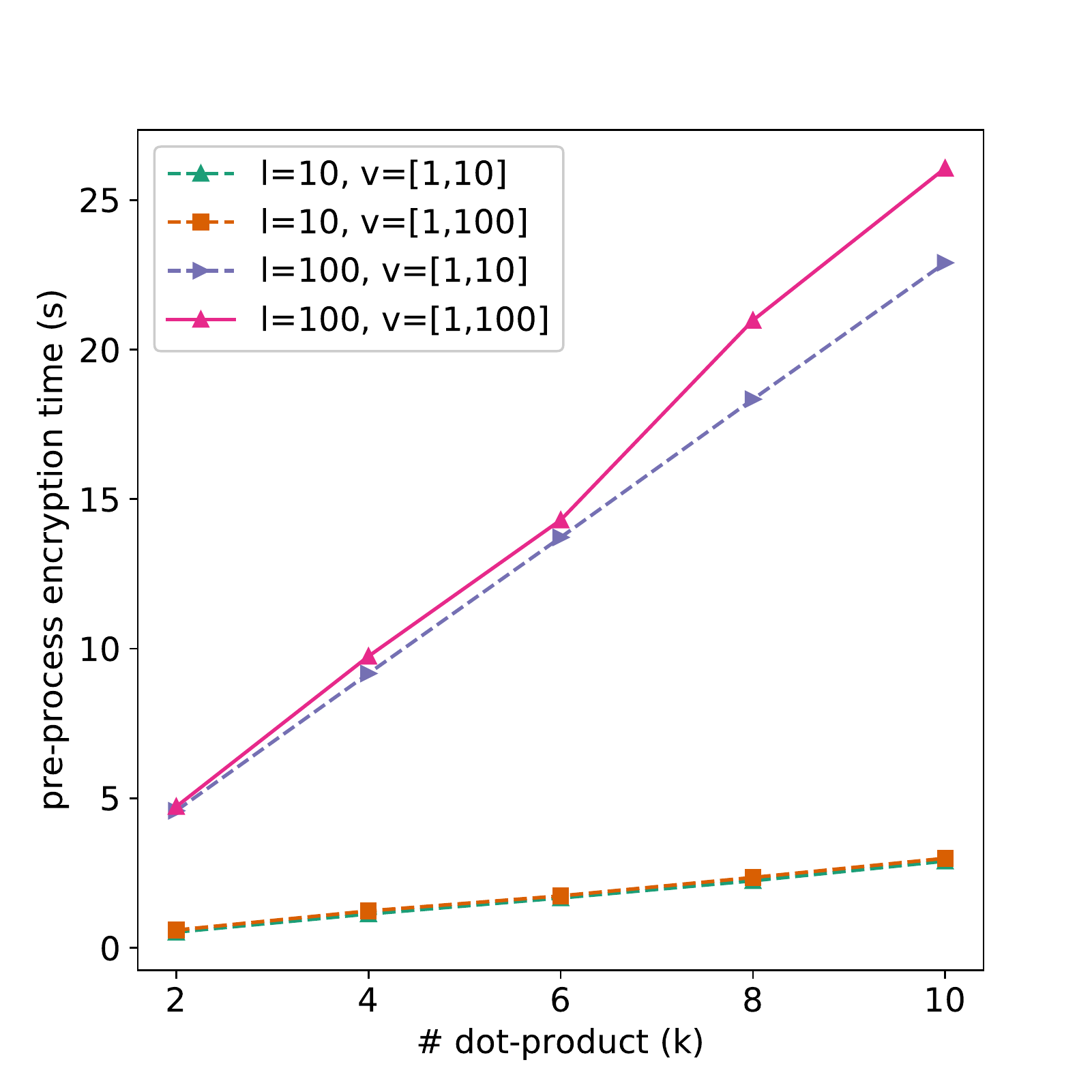}
        \label{fig:smc_ip:enc}
    }
    \hspace{-5mm}
    \subfloat[pre-processing for function key]{
        \includegraphics[scale=0.28]{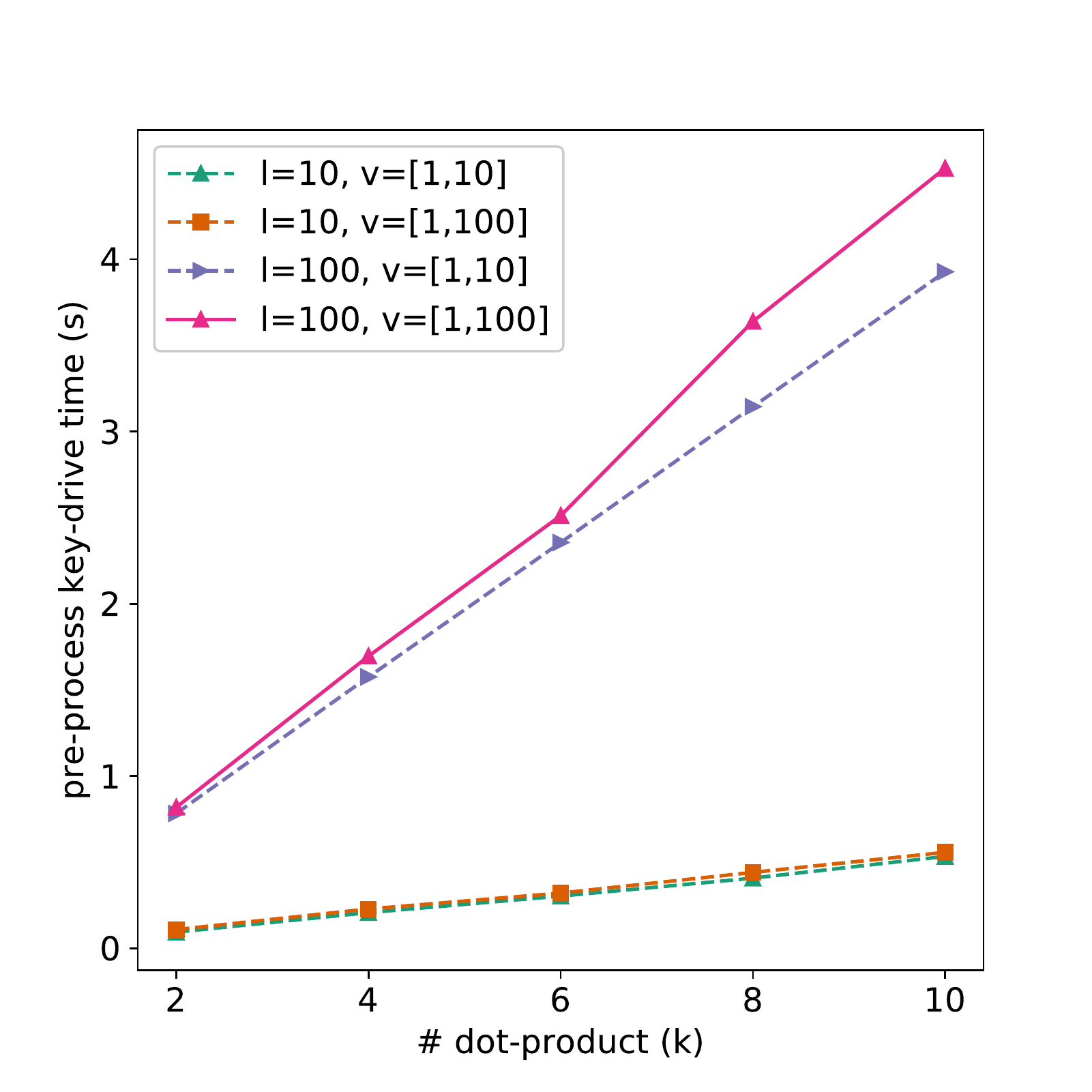}
        \label{fig:smc_ip:key}
    }
    \hspace{-5mm}
    \subfloat[secure dot-product computation]{
        \includegraphics[scale=0.28]{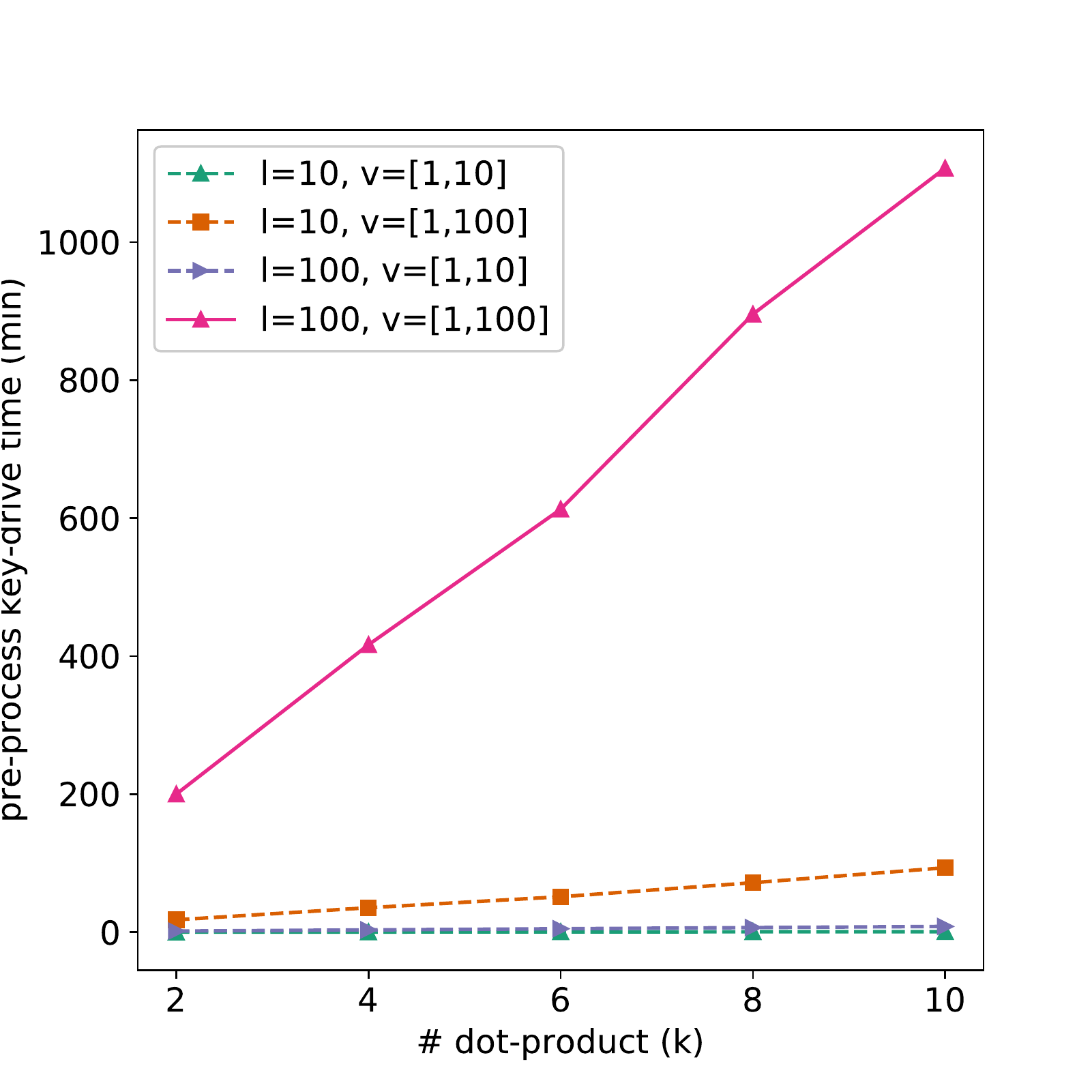}
        \label{fig:smc_ip:dec}
    }
    \hspace{-5mm}
    \subfloat[secure dot-product (parallelized)]{
        \includegraphics[scale=0.28]{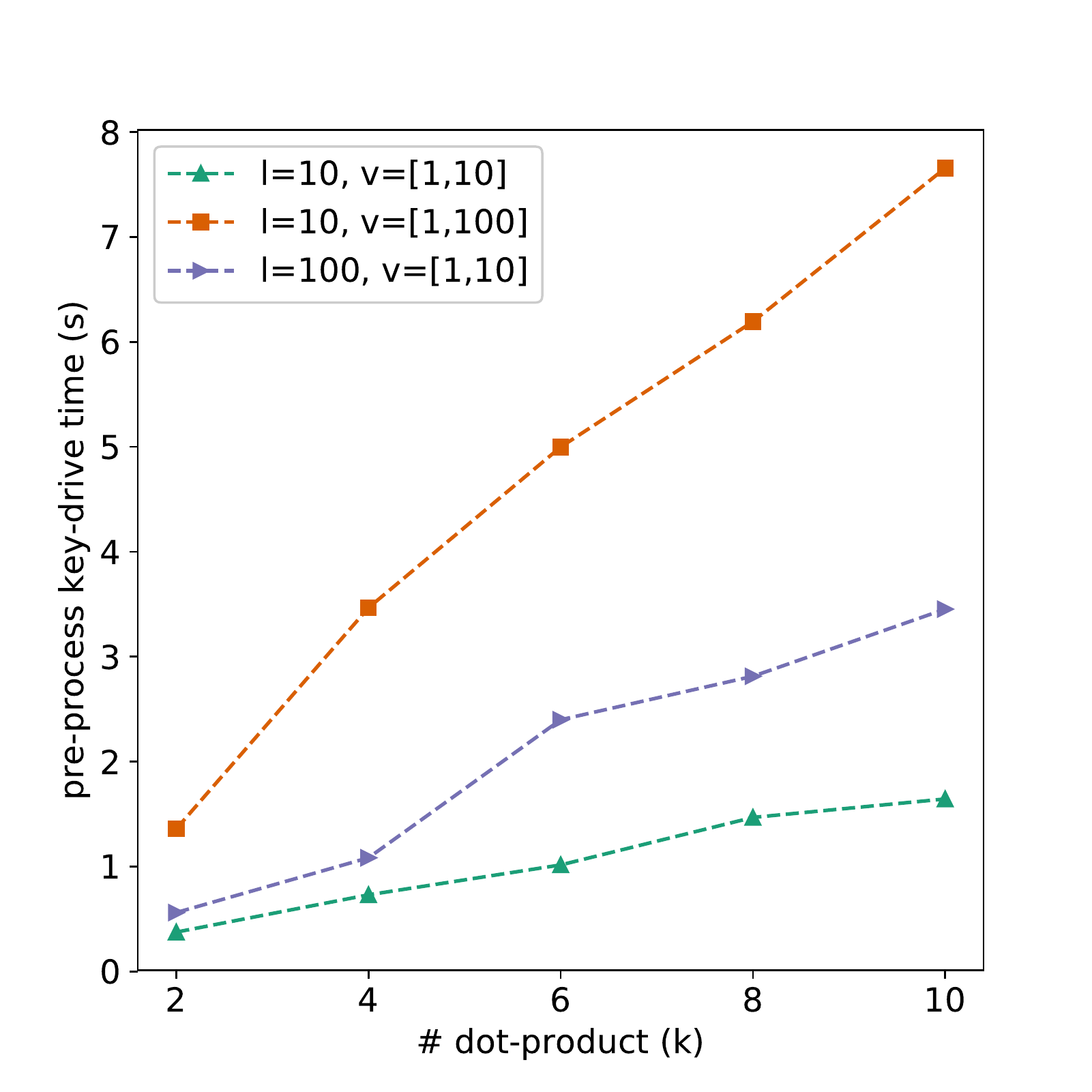}
        \label{fig:smc_ip:dec_parallel}
    }
    \vspace{-1mm}
    \caption{The time cost of dot-product in secure matrix computation scheme.} 
    \vspace{-6mm}
    \label{fig:smc_ip} 
\end{figure*}

\begin{figure}[t]
  \centering\includegraphics[scale=0.45]{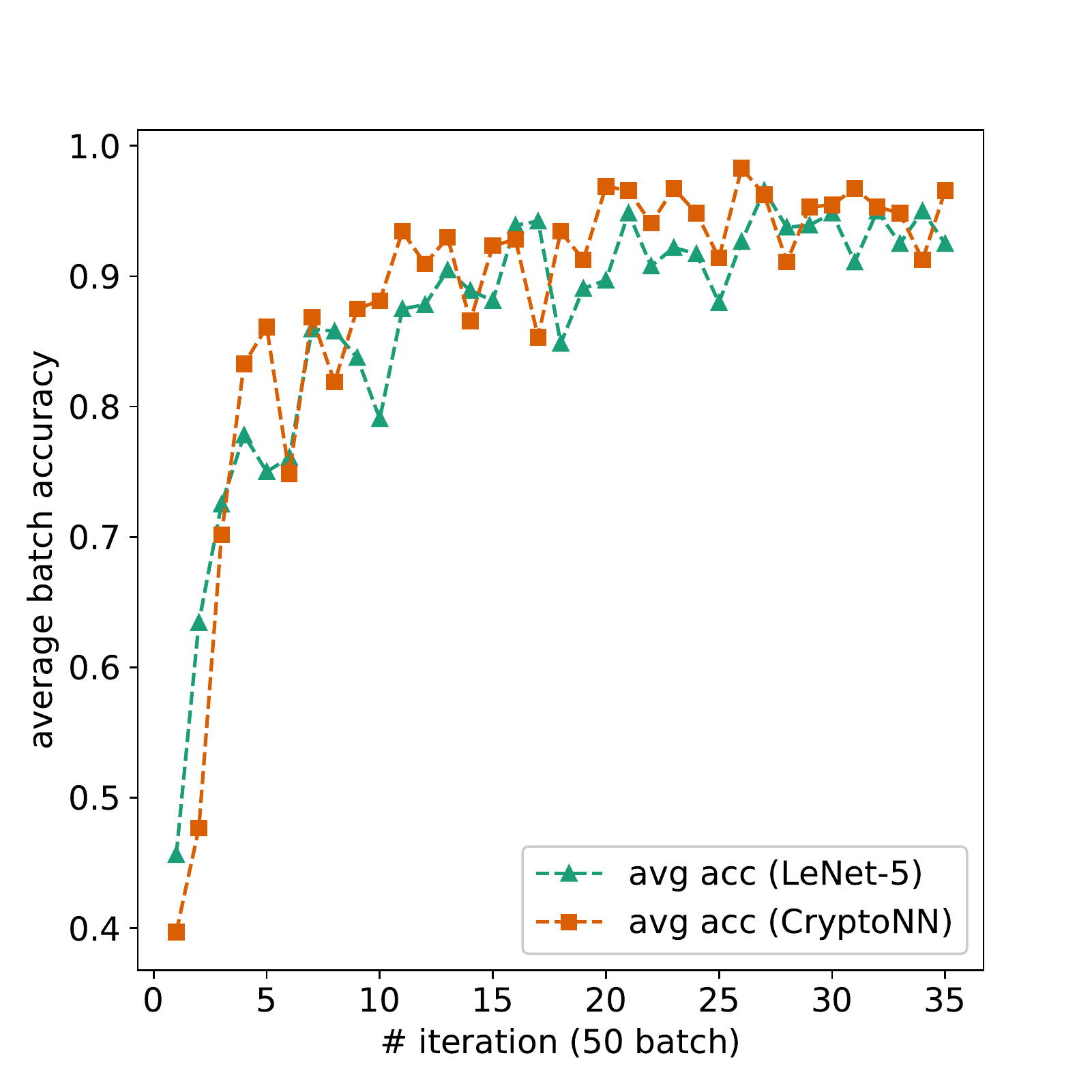}
  \vspace{-5mm}
  \caption{Comparison of average batch accuracy between original LeNet-5 and CryptoCNN.}
  \vspace{-5mm}
  \label{fig:avg_acc}
\end{figure}

\subsection{Performance Evaluation}
\subsubsection{Prototype implementation and test platform}
To verify the functionality and evaluate the performance, we implement a prototype of the CryptoNN framework.
The underlying crypto system is implemented using the Charm library \cite{charm13}.
Charm is a Python-based crypto toolkit and its underlying numerical calculations rely on GMP library.
The neural network model is implemented using Numpy
, a scientific computation library.

The test platform is a personal computer with Intel Core i7 processor, 16GB memory and macOS.
Note that the training process of the model only relies on the CPU.
In all the experiments, the security parameter is set to 256-bit.
The time measurement for program execution is based on the built-in \textit{time} package of python.

\subsubsection{Performance of secure matrix computation}

We evaluate the performance of the secure matrix computation scheme over the simulated encrypted data.
Due to consideration of page limitation, although the scheme supports several arithmetic computations, we only present the evaluation result of three typical arithmetic computations, namely, element-wise addition, element-wise product, and dot-product.
The execution times of those operations are depicted in \figurename \ref{fig:smc_add}, \figurename \ref{fig:smc_mul} and \figurename \ref{fig:smc_ip}.

\begin{table}[t]
    \centering
    \begin{threeparttable}
    \caption{The accuracy and training time}
    \label{table:val_acc}
    \begin{tabular}{llll}
        \toprule
        model     & epoch 1 (acc) & epoch 2 (acc) & training time  \\
        \midrule
        LeNet-5   & 93.04\% & 95.48\% & 4h\\
        CryptoCNN    & 93.12\% & 95.49\% & 57h\\
        \bottomrule
    \end{tabular}
    \end{threeparttable}
    \vspace{-5mm}
\end{table}

Specifically, the plaintext matrix is generated randomly in a specific range, as described in the legend of each figure. 
We measure the pre-processing times for encryption, generating function derived key, and the final secure computation, which are executed by the \textit{client}, \textit{authority}, and \textit{server}, respectively.
As depicted in \figurename \ref{fig:smc_add:enc} and \figurename \ref{fig:smc_mul:enc}, the encryption time is nearly linear to the element size in the matrix, where the x-axis represents the element size (k), and the y-axis denotes the processing time in ms.
Similarly, the execution time of pre-processing for the function derived key and element-wise computation also has a linear characteristic.
As shown in \figurename \ref{fig:smc_add:dec}, \figurename \ref{fig:smc_mul:dec}, and \figurename \ref{fig:smc_ip:dec}, the secure element-wise and dot-product computation are time-consuming.
To tackle that issue, we apply parallelization technique in the implementation.
As shown in \figurename \ref{fig:smc_add:dec_parallel}, \figurename \ref{fig:smc_mul:dec_parallel}, and \figurename \ref{fig:smc_ip:dec_parallel}, the parallelization technique makes secure computation more practical, where the execution time of secure dot-product can be reduced from near 90 mins to 8 seconds.

\noindent\textit{Communication overhead of key generation}. 
Suppose that for training a two-class classification NN model with $k$ units in the first hidden layer, the data set is $X_{m\times n}$, where $m$ is sample size and $n$ is feature size.
For each training iteration, the \textit{server} sends $k\times n \times |w|$ to the \textit{authority} and acquires the key with size of $k\times |\text{sk}|$, where $|w|$ and $|\text{sk}|$ are the size of one weight parameter and derivative key, respectively.

\subsubsection{Performance of CryptoNN}

We implement an instance of the CryptoNN framework, namely, CryptoCNN model, as described in Section \ref{sec:cryptonn:cnn}.
The model is built on the LeNet-5 architecture.
We train the model on the classic MNIST dataset \cite{lecun-mnisthandwrittendigit-2010} including a training set of 60000 examples, and a test set of 10000 examples, and then compare the performance on the test set.
Note that since the underlying functional encryption does not support floating point number computation and has lower efficiency on integers with longer lengths, we discard some precision on the parameter.
For instance, for a floating point parameter, we only keep two-decimal places approximately and then transfer the floating point number to the integer for the underlying crypto related computation.
Here, we only train two epochs using stochastic gradient descent method, where the batch size is set to 64.
The comparison of the accuracy and total training time is shown in \tablename$\;$\ref{table:val_acc}.
Also, the average batch accuracy of each model is depicted in \figurename \ref{fig:avg_acc}.
We can conclude that CryptoCNN has similar accuracy compared to the original model, while the training time is much longer than the original model due to the time-consuming cryptographic computations.

\section{Related Work}
\label{sec:related_work}

\noindent\textit{Functional Encryption}.
The concept of functional encryption was proposed by Amit Sahai and Brent Waters in \cite{sahai2005fuzzy}, and the formal definition was presented by Boneth et al. in \cite{boneh2011functional}.
As a new technique for public-key cryptography \cite{boneh2012functional}, functional encryption can be viewed as the abstraction of most of the existing encryption schemes such as attribute-based encryption \cite{ostrovsky2007attribute}, inner production function \cite{park2011inner}, and order-revealing encryption \cite{agrawal2004order}. 
Generally, unlike traditional encryption schemes, where decryption reveals all or nothing, in a functional encryption scheme, the decryption keys may reveal only partial information about the plaintext, for instance, a computed result of a specific function.

Several constructions of functional encryption has been proposed in the literature to deal with different functions such as inner-product \cite{kim2018function}, comparing the order \cite{boneh2011functional}, and access control \cite{lewko2010fully}.
On the other hand, researchers also have focused on the functional encryption construction using different existing techniques such as multi-party computation \cite{gorbunov2012functional} and multilinear maps \cite{lewi20165gen, carmer20175gen}, or using functional encryption to construct other cryptographic schemes such as indistinguishablility obfuscation \cite{garg2016candidate} and program obfuscation \cite{carmer20175gen}.
However, most of these schemes are in theoretical stages and not practical enough for applications.
Recently, more progressive approaches such as multi-input functional encryption \cite{goldwasser2014multi}, functional encryption using Intel SGX \cite{fisch2017iron} and practical implementation \cite{abdalla2015simple, kim2018function} have been proposed.
To achieve secure function computation, the work proposed in \cite{abdalla2015simple} is one of the underlying crypto systems in our proposed CryptoNN framework. 

\noindent\textit{Secure Machine Learning}.
The development of machine learning techniques, especially the emerging deep learning approaches \cite{lecun2015deep, krizhevsky2012imagenet} have enabled applications to be more intelligent than before.
The combination of machine learning and security study can be seen in two directions: (\romannumeral1) security issues in artificial intelligence, and (\romannumeral2) artificial intelligence for increasing the capability of existing security mechanisms \cite{biggio2018wild}.
A series of secure or privacy-preserving machine learning models have been proposed in \cite{gilad2016cryptonets, graepel2012ml, gonzalez2018supervised, wiesberg2018unsupervised,  hesamifard2017cryptodl, chabanne2017privacy, park2018efficient, jiang2018secure, mirhoseini2016cryptoml, shokri2015privacy, abadi2016deep, mohassel2017secureml, rouhani2018deepsecure,bost2015machine}.
Note that we only cover the discussion about the privacy-preserving machine learning models, rather than other security issues in AI such as adversary machine learning.

Most of the existing privacy-preserving approaches adopt either secure multi-party computation protocol (e.g., \cite{mohassel2017secureml, rouhani2018deepsecure}) or homophobic encryption (such as \cite{gilad2016cryptonets, graepel2012ml, gonzalez2018supervised, wiesberg2018unsupervised,  hesamifard2017cryptodl, chabanne2017privacy, park2018efficient, jiang2018secure}).
Neither of them supports training the machine learning model over encrypted data, except for the solution proposed in \cite{bost2015machine}, where the solution only supports limited machine learning model and also relies on the homomorphic encryption and the associated secure protocol designed.
To address the gap of training a machine learning model over encrypted data, our proposed CryptoNN focuses on training a secure deep learning model over the data encrypted using functional encryption scheme.

\vspace{-1mm}
\section{Conclusion}
\label{sec:conclusion}
Emerging neural networks based machine learning models such as deep learning and its variants have shown tremendous potential for many application domains that rely or use huge amounts of users' privacy sensitive data.
To tackle the serious privacy concerns of the collected data, several privacy-preserving machine learning approaches that use either secure multi-party computation or homomorphic encryption as the underlying mechanisms have been proposed in the literature.
In this paper, we have proposed a novel CryptoNN framework to support the training of neural networks over encrypted data using emerging functional encryption schemes.
The security analysis and performance evaluation show that CryptoNN achieves the privacy goal, as well as model accuracy.
One future direction is to explore the scalability of our CryptoNN framework for more complex datasets, and various other neural network models. Another future work is the use of different hardware support for CryptoNN.

\section*{Acknowledgment}
This research work has been supported by the National Science Foundation grant DGE-1438809.

\bibliographystyle{IEEEtran} 
\bibliography{reference.bib}

\end{document}